\documentclass[12pt]{article}

\usepackage{amssymb,amsmath,comment,mathtools}
\usepackage{braket}
\usepackage{bm}

\usepackage[dvipdfmx]{graphicx}
\usepackage{subfigmat}
\usepackage{here}
\usepackage{booktabs}
\graphicspath{{./Figures/}}

\usepackage[dvipdfmx]{xcolor}

\usepackage[
  dvipdfmx,
  bookmarksnumbered=true,
  linktocpage=true,
  setpagesize=false
]{hyperref}

\hypersetup{
  colorlinks=false,           
  citebordercolor=green,
  linkbordercolor=red,
  urlbordercolor=cyan,
}

\usepackage{csquotes}
\usepackage[
  backend=biber,
  style=phys,
  biblabel=brackets,
  sorting=none,
  articletitle=true,
  giveninits=true,
  maxbibnames=99,
  doi=false,
  url=true,
  eprint=true
]{biblatex}
\DeclareFieldFormat{eprint:arxiv}{%
  \mbox{arXiv\addcolon\addnbspace}%
  \href{https://arxiv.org/abs/#1}{\mbox{\nolinkurl{#1}}}%
}
\usepackage{etoolbox}
\AtBeginBibliography{%
  \small
  \sloppy
  \hyphenpenalty=50
  \interlinepenalty=10
}
\newcommand{\ba}{\begin{align}}
\newcommand{\ea}{\end{align}}

\renewcommand{\thefootnote}{\fnsymbol{footnote}}

\begin{document}
\begin{flushright}
{\small
NITEP 280}
\end{flushright}

\vspace*{1.0cm}

\begin{center}
\baselineskip 20pt 
{\Large\bf 
Correlation Between Proton Decay Channels and the Axion Mass in an Extended SU(5) GUT}
\vspace{1cm}

{\large 
Naoyuki Haba${}^{a,b}$, Keisuke Nagano${}^{a}$, Yasuhiro Shimizu${}^{a,b}$ \\
and Toshifumi Yamada${}^{c}$
} \vspace{.5cm}

{\baselineskip 20pt \it
${}^{a}$Department of Physics, Osaka Metropolitan University, Osaka 558-8585, Japan \\
${}^{b}$Nambu Yoichiro Institute of Theoretical and Experimental Physics (NITEP),
Osaka Metropolitan University, Osaka 558-8585, Japan\\
${}^{c}$Institute for Mathematical Informatics, Meiji Gakuin University, Yokohama 244-8539, Japan
}

\vspace{.5cm}

\vspace{1.5cm} {\bf Abstract} \end{center}

We study a renormalizable SU(5) grand unified theory (GUT) supplemented by a 45-dimensional Higgs field and a DFSZ axion sector, imposing a Georgi--Jarlskog flavor structure at the unification scale. 
We perform a one-loop gauge coupling unification analysis, explicitly including the threshold masses of the light multiplets arising from the 45-dimensional Higgs field. 
This analysis identifies the viable region in the $(M_{\mathrm{GUT}},M_{S_1})$ parameter space. 
Through the relation between the unification and PQ scales, this region yields correlated predictions for the QCD axion mass.
The Georgi--Jarlskog assumption substantially reduces the flavor ambiguity of the dimension-six baryon-violating operators, enabling robust constraints and predictions not only for antineutrino modes but also for charged-lepton proton decay modes such as $p \to e^+ \pi^0$. 
We present the combined implications for proton decay and axion searches, showing how the GUT-selected parameter region maps onto the axion mass, the axion-photon coupling, and the axion-induced EDM coupling.

\thispagestyle{empty}

\newpage
\renewcommand{\thefootnote}{\arabic{footnote}}
\setcounter{footnote}{0}
\baselineskip 18pt

\section{Introduction}

\indent
The Standard Model (SM) has been remarkably successful in describing elementary particle interactions up to the electroweak scale. 
Nevertheless, it leaves several theoretical questions unanswered. 
Two issues are particularly relevant to the present work: the apparent unification of the gauge interactions and the strong CP problem in QCD. 
The former suggests a grand-unified description of the SM gauge interactions, whereas the latter calls for an additional dynamical mechanism beyond the SM.

Grand Unified Theories (GUT) provide a simple framework in which the SM gauge interactions are embedded into a single gauge group \cite{PatiSalam1973,PatiSalam1974,GeorgiGlashow1974,GeorgiQuinnWeinberg1974,Georgi1975GaugeTheories,FritzschMinkowski1975}.
Among them, the SU(5) GUT is one of the most economical possibilities. 
However, the minimal Georgi--Glashow model does not yield fully realistic phenomenology \cite{GeorgiGlashow1974}.
In particular, the minimal Higgs sector predicts unrealistic charged-fermion mass relations. 
A well-known way to overcome this difficulty is to introduce a 45-dimensional Higgs representation, as in the Georgi--Jarlskog mechanism \cite{GeorgiJarlskog1979}. 
The scalar components of the 45-dimensional Higgs field can also modify the threshold corrections relevant to gauge coupling unification and affect proton decay predictions. 
Thus, non-minimal SU(5) models with a 45-dimensional Higgs field provide a motivated framework for studying fermion masses, gauge coupling unification, and proton decay within a common setup.

The strong CP problem can be solved by the Peccei--Quinn (PQ) mechanism, which predicts the QCD axion as a pseudo-Nambu--Goldstone boson \cite{Peccei:1977hh,Peccei:1977ur}. 
The axion can also constitute a viable dark-matter candidate. 
In the present context, the DFSZ axion model is particularly natural because it realizes the PQ symmetry through an extended Higgs sector and the ordinary SM fermions, without introducing additional heavy vector-like quarks \cite{Dine:1981rt,Zhitnitsky:1980tq}. 
This feature makes the DFSZ framework compatible with non-minimal SU(5) constructions, where additional Higgs representations are already required for realistic fermion mass relations. 
The DFSZ axion therefore provides a natural link between the strong CP problem and the Higgs structure of grand unification.

This link becomes especially predictive when the PQ sector is tied to the grand-unified symmetry breaking sector. 
In generic PQ scenarios, the symmetry breaking scale remains largely independent, and low-energy axion phenomenology does not strongly restrict the ultraviolet theory. 
By contrast, in grand-unified realizations, the axion decay constant can become correlated with the GUT scale and the heavy-field spectrum. 
SU(5) axion models have been studied since the early invisible-axion construction~\cite{Wise:1981ry}. 
Previous studies of SU(5) axion models have shown that different embeddings can lead to similar low-energy axion phenomenology once unphysical PQ charge ambiguities are properly identified~\cite{Quevillon:2020}.
Therefore, low-energy axion observables alone do not uniquely determine the underlying unified model. 
A realistic 45-dimensional Higgs sector offers an additional handle, since scalar thresholds, gauge coupling unification, proton decay, and axion phenomenology become mutually correlated.

In this work, we study an axion-extended renormalizable SU(5) GUT with a 45-dimensional Higgs field. 
We specify the field content and PQ charge assignments, derive the physical axion mode, and obtain the relation between the axion decay constant and the vacuum structure responsible for PQ breaking. 
We then analyze gauge coupling unification in the presence of hierarchical scalar thresholds from the 45-dimensional Higgs field. 
We impose proton decay constraints and identify the viable parameter region. 
In this region, we evaluate the resulting axion predictions and clarify how the axion mass and couplings correlate with the GUT scale, scalar spectrum, and proton decay observables.

This paper is organized as follows. 
In Sec.\,\ref{sec:su(5) GUT with 45}, we introduce the renormalizable SU(5) model with a 45-dimensional Higgs representation. 
In Sec.\,\ref{sec:DFSZ}, we implement the DFSZ mechanism and derive the relation among the PQ scale, the GUT scale, and the axion mass. 
In Sec.\,\ref{sec:GCU}, we analyze gauge coupling unification with hierarchical scalar thresholds. 
In Sec.\,\ref{sec:proton decay}, we impose proton decay constraints and identify the allowed parameter region. 
In Sec.\,\ref{sec:axion}, we discuss axion phenomenology and compare the predictions with projected experimental sensitivities. 
Section\,\ref{sec:conclusion} summarizes our results.

\section{SU(5) GUT with the 45-dimensional Higgs}
\label{sec:su(5) GUT with 45}

In this section, we consider a renormalizable SU(5) GUT without assuming supersymmetry. The fermions in the SM are embedded into the $\bm{\bar{5}}$ and $\bm{10}$ representations of SU(5) as follows:
\begin{align}
\bm{\bar{5}}: \overline{\psi}^a_i
=
\begin{pmatrix}
d_{Ri}^{Ca} &
\epsilon_{\alpha\beta}\, 
\ell_{Li}^{\, \beta} 
\end{pmatrix},~~~~
\bm{10}:\chi_{i\,ab} =
\frac{1}{\sqrt{2}}
\begin{pmatrix}
\epsilon^{abc}\,
u^{C}_{Ri\,c} &
q_{Li}^{\, \hat{a}\beta}
\\[1mm]
-q_{Li}^{\, b\alpha} &
\epsilon^{\alpha\beta}\, 
e_{Ri}^{\, c}
\end{pmatrix}.
\end{align}
Here, $a,b,c$ are $SU(3)_C$ indices, $\alpha,\beta$ are $SU(2)_L$ indices, and $i,j$ are generation indices.
The scalar sector contains an adjoint scalar field $\Sigma \sim 24$, which breaks $SU(5)$ down to the SM gauge group. 
The scalar sector also contains the Higgs multiplets $\Phi_5 \sim 5$, $\Phi_5^\prime \sim 5$, and $\Phi_{45} \sim 45$. 
In contrast to the minimal Georgi--Glashow model, we take $\Sigma$ to be a complex scalar in the adjoint representation. 
This assumption is necessary in the present PQ realization, because $24_H$ carries a nonzero PQ charge and the SM-singlet component $\Sigma_1$ contains the axion phase, as shown in Eq.~\eqref{eq:Sigma_phase}. 
The usual real adjoint corresponds to the case in which the adjoint field is Hermitian and carries no continuous PQ phase.

The $\mathbf{5}$-dimensional Higgs multiplet decomposes into an electroweak doublet and a color-triplet scalar. 
Since these two components must lie at vastly different mass scales, the model faces the usual doublet--triplet splitting problem. 
We do not address this problem in the present work.
Instead, we assume that the color-triplet component in $\Phi_5$ is decoupled and neglect its contribution to proton decay. 
By contrast, the light color anti-triplet state dominated by the $\Phi_{45}$ component is treated explicitly in Sec.\,\ref{sec:proton decay}.

In the minimal renormalizable SU(5) GUT, fermion masses arise from Yukawa interactions involving the Higgs field in the $\mathbf{5}$ representation. 
This minimal Higgs sector, however, leads to the unification-scale relation
$Y_d = Y_e^T$, which gives phenomenologically unacceptable mass relations between down-type quarks and charged leptons. 
Thus, we consider an extended scalar sector within a renormalizable SU(5) framework. 
Table~\ref{tab:scalar_decomposition} summarizes the decomposition of the scalar multiplets under the SM gauge group.

\begin{table}[h]
\centering
\caption{The decomposition of the scalar fields $\Sigma$, $\Phi_5$, and $\Phi_{45}$ under the SM gauge groups.}
\label{tab:scalar_decomposition}
\vspace{4mm}
\resizebox{\textwidth}{!}{
\begin{tabular}{|l|l|c|c|c|c||l|l|c|c|c|c|}
\hline
Field & SU(5) & Field & $SU(3)_C$ & $SU(2)_L$ & $U(1)_Y$
&
Field & SU(5) & Field & $SU(3)_C$ & $SU(2)_L$ & $U(1)_Y$
\\
\hline
$\Sigma$ & $\mathbf{24}$
& $\Sigma_1$ & $\mathbf{1}$ & $\mathbf{1}$ & $0$
&
$\Phi_{45}$ & $\mathbf{45}$
& $\widetilde{S}_1$ & $\overline{\mathbf{3}}$ & $\mathbf{1}$ & $4/3$
\\
&
& $\Sigma_3$ & $\mathbf{1}$ & $\mathbf{3}$ & $0$
&
&
& $R_2^*$ & $\overline{\mathbf{3}}$ & $\overline{\mathbf{2}}$ & $-7/6$
\\
&
& $\Sigma_G$ & $\mathbf{3}$ & $\overline{\mathbf{2}}$ & $-5/6$
&
&
& $S_3^*$ & $\mathbf{3}$ & $\mathbf{3}$ & $-1/3$
\\
&
& $\Sigma_G^*$ & $\overline{\mathbf{3}}$ & $\mathbf{2}$ & $5/6$
&
&
& $S_6^*$ & $\overline{\mathbf{6}}$ & $\mathbf{1}$ & $-1/3$
\\
&
& $\Sigma_8$ & $\mathbf{8}$ & $\mathbf{1}$ & $0$
&
&
& $S_8$ & $\mathbf{8}$ & $\mathbf{2}$ & $1/2$
\\
$\Phi_5$ & $\mathbf{5}$
& $H^{(5)}$ & $\mathbf{1}$ & $\mathbf{2}$ & $1/2$
&
&
& $H^{(45)}$ & $\mathbf{1}$ & $\mathbf{2}$ & $1/2$
\\
&
& $S_1^{(5)*}$ & $\mathbf{3}$ & $\mathbf{1}$ & $-1/3$
&
&
& $S_1^{(45)*}$ & $\mathbf{3}$ & $\mathbf{1}$ & $-1/3$
\\
$\Phi_5^\prime$ & $\mathbf{5}$
& $H^{\prime(5)}$ & $\mathbf{1}$ & $\mathbf{2}$ & $1/2$
&
&
&
&
&
&
\\
&
& $S_1^{\prime(5)*}$ & $\mathbf{3}$ & $\mathbf{1}$ & $-1/3$
&
&
&
&
&
&
\\
\hline
\end{tabular}
}
\end{table}
The extended scalar sector contains the additional five-dimensional multiplet $\Phi_5^\prime$ and the 45-dimensional multiplet $\Phi_{45}$. 
The multiplet $\Phi_5^\prime$ realizes the PQ charge assignment and the DFSZ-type Higgs structure. 
The 45-dimensional multiplet $\Phi_{45}$ modifies the minimal relation $Y_d = Y_e^T$, thereby allowing realistic mass relations between down-type quarks and charged leptons.

After the vacuum expectation value of the SM-singlet component $\Sigma_1\subset\Sigma$ breaks SU(5) down to the SM gauge group,
we focus on the electroweak Higgs doublets $H^{(5)}\subset\Phi_5$, $H^{(45)}\subset\Phi_{45}$, and $H^{\prime(5)}\subset\Phi_5^\prime$.
These fields, together with $\Sigma_1$, acquire the vacuum expectation values $v_1$, $v_2$, $v_3$, and $v_\Sigma$, respectively.
After omitting radial fluctuations, the neutral components of the Higgs fields are written as
\begin{align}
	H^{(5)}_0(x)
	\simeq&~\frac{v_1}{\sqrt{2}}
	\exp\biggl[i\,\frac{a_1(x)}{v_1}\biggr]
	= \frac{v_1}{\sqrt{2}}
	\exp\Bigl[i\bigl(q\,\hat{a}_Z+\mathrm{PQ}_1\,\hat a\bigr)\Bigr],
	\label{eq:Goldstone1}\\
	H^{(45)}_0(x)
	\simeq&~\frac{v_2}{\sqrt{2}}
	\exp\biggl[i\,\frac{a_2(x)}{v_2}\biggr]
	= \frac{v_2}{\sqrt{2}}
	\exp\Bigl[i\bigl(q\,\hat{a}_Z+\mathrm{PQ}_2\,\hat a\bigr)\Bigr],\\
	H^{\prime(5)}_0(x)
	\simeq&~\frac{v_3}{\sqrt{2}}
	\exp\biggl[i\,\frac{a_3(x)}{v_3}\biggr]
	= \frac{v_3}{\sqrt{2}}
	\exp\Bigl[i\bigl(q\,\hat{a}_Z+\mathrm{PQ}_3\,\hat a\bigr)\Bigr],
	\label{eq:Goldstone2} \\
	\Sigma_1(x)
	\simeq&~\frac{v_\Sigma}{\sqrt{2}}
	\exp\biggl[i\,\frac{a_\Sigma(x)}{v_\Sigma}\biggr]
	= \frac{v_\Sigma}{\sqrt{2}}
	\exp\bigl[i\mathrm{PQ}_\Sigma\,\hat a\bigr].
	\label{eq:Sigma_phase}
\end{align}
Here, $\hat{a}_Z$ denotes the neutral Goldstone boson eaten by the $Z$ boson, while $\hat a$ denotes the axion direction.

In this model, we impose a global $U(1)_{\rm PQ}$ symmetry with the following charge assignment:
\begin{center}
\begin{tabular}{r | c c c c c c }
Field
& $\bar{5}$
& $10$
& $5'_H$
& $5_H$
& $45_H$
& $24_H$ \\
\hline
PQ charge
& $\alpha$
& $\beta$
& $-2\beta$
& $\alpha+\beta$
& $\alpha+\beta$
& $-(\alpha+3\beta)/2$
\end{tabular}
\end{center}
This charge assignment controls which Higgs multiplets can couple to the fermions.
In particular, the SU(5) and $U(1)_{\rm PQ}$ invariant Yukawa interactions are given by
\begin{align}
-\mathcal{L}_Y
  &=
  \frac{1}{8}\, (Y_5^U)_{ij}\epsilon^{abcde}\,
    \chi_{i\,ab}\,\chi_{j\,cd}\,(5_H^\prime)_e
  + (Y_5^D)_{ij}
    \chi_{i\,ab}\, \overline{\psi}^a_j \,(5_H^\dagger)^b
  + \frac{1}{2}\, (Y_{45}^D)_{ij}
    \chi_{i\,ab}\,\overline{\psi}^c_j({45}_H^\dagger)_{ab}^c
  + \mathrm{h.c.}.
    \label{eq:GUTYukawa}
\end{align}
After SU(5) is spontaneously broken, the electroweak doublet contained in $5'_H$ generates the up-type quark masses, whereas those contained in $5_H$ and $45_H$ generate the down-type quark and charged-lepton masses after electroweak symmetry breaking. 

The PQ charge assignment introduced above determines the PQ invariant combinations of the scalar fields. 
In particular, it restricts the interactions involving the adjoint scalar $24_H$, so that the scalar potential contains the following $24_H$-dependent terms:
\begin{align}
V_{24}
=&~
\Phi_5^{\prime\dagger}\,\Sigma^2
(\lambda_1\,\Phi_5+\lambda_2\,\Phi_{45})
+\lambda_3\,\Phi_5^{\prime\dagger}\,\Phi_5\,\mathrm{Tr}[\Sigma^2]
+\mathrm{h.c.}.
\label{eq:DFSZterms}
\end{align}
Moreover, the fields $5_H$ and $45_H$ alone also allow the quartic interaction
\begin{align}
V_{5~45}
=&~
\lambda\,\Phi_5^\dagger\,\Phi_{45}\,\Phi_{45}^\dagger\,\Phi_5
+\mathrm{h.c.}.
\label{eq:quartic545}
\end{align}
After SU(5) is spontaneously broken by the vacuum expectation value of $\Sigma_1$, these scalar-potential terms can be expressed in terms of $H^{(5)}$, $H^{(45)}$, $H^{\prime(5)}$, and the SM-singlet component $\Sigma_1$.

\section{The DFSZ mechanism in an extended SU(5) GUT}
\label{sec:DFSZ}
Having introduced the PQ charge assignment and the corresponding scalar interactions in previous section, 
we now discuss how these ingredients determine the axion sector of the extended SU(5) GUT.

The PQ charge assignment introduced above restricts the scalar interactions and fixes the structure of the terms that control the relative phases of these fields.
The relevant part of the scalar potential can then be written as
\begin{align}
	V_{\rm phase}
	=&~
	(
	m_{12}^2 + \lambda H_1^\dagger H_2
	) H_1^\dagger H_2
	+
	H_3^\dagger \Sigma_0^2
	(
	\lambda_{a1} H_1 + \lambda_{a2} H_2
	)
	+ \mathrm{h.c.}.
	\label{eq:scalar-potential-sm}
\end{align}
Substituting the phase parametrization of the neutral scalar components into the phase-sensitive terms in Eq.~\eqref{eq:scalar-potential-sm}, we obtain
\begin{align}
H_1^\dagger H_2 \Bigr|_{\rho=0}
=&~
\frac{v_1 v_2}{2}
\exp\biggl[-i\frac{a_1}{v_1}+i\frac{a_2}{v_2}\biggr]
=
\frac{v_1 v_2}{2}
\exp\Bigl[i\bigl(\mathrm{PQ}_2-\mathrm{PQ}_1\bigr)\hat a\Bigr],\\
H_3^\dagger \Sigma_0^2 H_i \Bigr|_{\rho=0}
=&~
\frac{v_3 v_\Sigma^2 v_i}{2\sqrt{2}}
\exp\biggl[-i\frac{a_3}{v_3}+2i\frac{a_\Sigma}{v_\Sigma}+i\frac{a_i}{v_i}\biggr]
\nonumber\\
=&~
\frac{v_3 v_\Sigma^2 v_i}{2\sqrt{2}}
\exp\Bigl[i\bigl(\mathrm{PQ}_i+2\mathrm{PQ}_\Sigma-\mathrm{PQ}_3\bigr)\hat a\Bigr],
\qquad (i=1,2).
\end{align}
Since the axion corresponds to the Nambu--Goldstone mode of the spontaneously broken PQ symmetry, the scalar potential must not generate an axion-dependent term along this direction. 
Thus, the phase combinations in Eq.~\eqref{eq:scalar-potential-sm} impose
\begin{align}
-\frac{a_1}{v_1}+\frac{a_2}{v_2}
=&~0
\qquad
\Longleftrightarrow
\qquad
\mathrm{PQ}_1=\mathrm{PQ}_2,
\\
-\frac{a_3}{v_3}
+2\frac{a_\Sigma}{v_\Sigma}
+\frac{a_i}{v_i}
=&~0
\qquad
\Longleftrightarrow
\qquad
\mathrm{PQ}_i+2\mathrm{PQ}_\Sigma-\mathrm{PQ}_3=0,
\qquad i=1,2.
\end{align}

The above phase parametrization also allows us to identify the physical axion mode from the kinetic terms of the CP-odd scalars. Substituting \eqref{eq:Goldstone1}-\eqref{eq:Goldstone2} into the kinetic terms, one obtains
\begin{align}
	\frac{1}{2}\partial_\mu a_1 \partial^\mu a_1 =&~ \frac{1}{2} v_1^2 \left( q^2 \partial_\mu \hat{a}_Z \partial^\mu \hat{a}_Z + 2 q \, \text{PQ}_1 \,  \partial_\mu \hat{a}_Z \partial^\mu \hat{a} + \text{PQ}_1^2 \, \partial_\mu \hat{a} \partial^\mu \hat{a} \right),\\
	\frac{1}{2}\partial_\mu a_2 \partial^\mu a_2 =&~ \frac{1}{2} v_2^2 \left( q^2 \partial_\mu \hat{a}_Z \partial^\mu \hat{a}_Z + 2 q \, \text{PQ}_2 \, \partial_\mu \hat{a}_Z \partial^\mu \hat{a} + \text{PQ}_2^2 \, \partial_\mu \hat{a} \partial^\mu \hat{a} \right),\\
	\frac{1}{2}\partial_\mu a_3 \partial^\mu a_3 =&~ \frac{1}{2} v_3^2 \left( q^2 \partial_\mu \hat{a}_Z \partial^\mu \hat{a}_Z + 2 q \, \text{PQ}_3 \,  \partial_\mu \hat{a}_Z \partial^\mu \hat{a} + \text{PQ}_3^2  \, \partial_\mu \hat{a} \partial^\mu \hat{a} \right),\\
	\frac{1}{2}\partial_\mu a_\Sigma \partial^\mu a_\Sigma =&~ \frac{1}{2} v_\Sigma^2 \, \text{PQ}_\Sigma^2 \, \partial_\mu \hat{a} \partial^\mu \hat{a} .
\end{align}
The orthogonality of the physical axion to the neutral Goldstone boson eaten by the $Z$ boson requires the mixed term
$\partial_\mu \hat a_Z \partial^\mu \hat a$
to vanish. We therefore obtain
\begin{align}
v_1^2 \,\mathrm{PQ}_1 + v_2^2 \,\mathrm{PQ}_2 + v_3^2 \,\mathrm{PQ}_3
=&~ 0.
\label{eq:axionorth}
\end{align}

The remaining $\partial_\mu \hat a \partial^\mu \hat a$ terms determine the normalization of the physical axion direction. We thus define
\begin{align}
v_1^2 \,\mathrm{PQ}_1^2
+ v_2^2 \,\mathrm{PQ}_2^2
+ v_3^2 \,\mathrm{PQ}_3^2
+ v_\Sigma^2 \,\mathrm{PQ}_\Sigma^2
=&~ n_a^2,
\label{eq:naDef}
\end{align}
where $n_a$ is the normalization factor of the CP-odd phase associated with the physical axion. The canonically normalized axion field is then introduced as
\begin{align}
\hat a
=&~ \frac{a}{n_a},
&
n
=&~ \frac{n_a}{\mathrm{PQ}_1}.
\label{eq:ahatDef}
\end{align}
With this definition, the phase components of $H_1^0$ and $H_2^0$ can be written as
\begin{align}
\frac{a_1}{v_1}
=&~ \frac{a_2}{v_2}
= q\,\hat a_Z + \frac{a}{n}.
\label{eq:a12overv}
\end{align}

Using $\mathrm{PQ}_1=\mathrm{PQ}_2$ together with Eqs.~\eqref{eq:Goldstone2} and \eqref{eq:axionorth}, and defining the electroweak vacuum expectation value by $v^2 \equiv v_1^2+v_2^2+v_3^2$, one finds
\begin{align}
\mathrm{PQ}_3
=&~ -\frac{v_1^2+v_2^2}{v_3^2}\,\mathrm{PQ}_1,
\label{eq:PQ3sol}
\\
\mathrm{PQ}_\Sigma
=&~ \frac{\mathrm{PQ}_3-\mathrm{PQ}_1}{2}
= -\frac{v_1^2+v_2^2+v_3^2}{2v_3^2}\,\mathrm{PQ}_1
= -\frac{v^2}{2v_3^2}\,\mathrm{PQ}_1.
\label{eq:PQSsol}
\end{align}
Substituting these expressions back into the phase decomposition, we obtain
\begin{align}
\frac{a_3}{v_3}
=&~ q\,\hat a_Z + \mathrm{PQ}_3\,\hat a
= q\,\hat a_Z
-\left(\frac{v_1^2+v_2^2}{v_3^2}\right)\mathrm{PQ}_1\,\hat a
= q\,\hat a_Z
-\left(\frac{v_1^2+v_2^2}{v_3^2}\right)\frac{a}{n},
\label{eq:a3overv3}
\\
\frac{a_\Sigma}{v_\Sigma}
=&~ \mathrm{PQ}_\Sigma\,\hat a
= -\left(\frac{v^2}{2v_3^2}\right)\mathrm{PQ}_1\,\hat a
= -\left(\frac{v^2}{2v_3^2}\right)\frac{a}{n}.
\label{eq:aSigovervSig}
\end{align}
Here we have used the fact that the electroweak-singlet PQ breaking field does not contribute to the neutral Goldstone mode $q\,\hat a_Z$, since it is neutral under the electroweak gauge group.

Combining Eqs.~\eqref{eq:naDef}, \eqref{eq:PQ3sol}, and \eqref{eq:PQSsol}, $n$ can be expressed as
\begin{align}
n
=&~ \frac{n_a}{\mathrm{PQ}_1}
= \sqrt{
\frac{(v_1^2+v_2^2)\,v^2}{v_3^2}
+\frac{v_\Sigma^2\,v^4}{4v_3^4}
}
\simeq \frac{v^2\,v_\Sigma}{2v_3^2},
\qquad
(v_1,v_2,v_3 \ll v_\Sigma).
\label{eq:nDef}
\end{align}
The last approximation follows from the hierarchy $v_1,v_2,v_3 \ll v_\Sigma$, for which the second term inside the square root dominates parametrically.

After removing the unphysical $q\,\hat a_Z$ component, the axion-dependent parts of the neutral scalar fields are given by
\begin{align}
H^{(5)}_0
	\simeq &~\frac{v_1}{\sqrt{2}}
\exp\left(i\frac{a}{n}\right), &
H^{(45)}_0
	\simeq &~ \frac{v_2}{\sqrt{2}}
\exp\left(i\frac{a}{n}\right), \nonumber\\
H^{\prime(5)}_0
	\simeq &~\frac{v_3}{\sqrt{2}}
\exp\left[
-i\left(\frac{v_1^2+v_2^2}{v_3^2}\right)\frac{a}{n}
\right], &
\Sigma_1
\simeq &~\frac{v_\Sigma}{\sqrt{2}}
\exp\left[
-i\left(\frac{v^2}{2v_3^2}\right)\frac{a}{n}
\right].
\label{eq:axionrep}
\end{align}
Substituting these phases into the Standard Model fermion mass terms, one obtains
\begin{align}
\mathcal{L}_Y
=&~ M_u\,\overline{u_R}\,u_L\,
\exp\left[
-i\left(\frac{v_1^2+v_2^2}{v_3^2}\right)\frac{a}{n}
\right]
+ M_d\,\overline{d_R}\,d_L\,
\exp\left(-i\frac{a}{n}\right) \nonumber\\
&~+ M_e\,\overline{e_R}\,e_L\,
\exp\left(-i\frac{a}{n}\right)
+ \text{h.c.}
\label{eq:YukPhase}
\end{align}
Thus, the axion-dependent phases in the Yukawa sector arise directly from the phases of the Higgs vacuum expectation values.

These phases can be removed by the chiral field redefinitions
\begin{align}
u_{L/R}
\to&~ \exp\left[
\pm i\left(\frac{v_1^2+v_2^2}{2v_3^2}\right)\frac{a}{n}
\right] u_{L/R},
&
d_{L/R}
\to&~ \exp\left(\pm i\frac{a}{2n}\right)d_{L/R}, &
e_{L/R}
\to&~ \exp\left(\pm i\frac{a}{2n}\right)e_{L/R}.
\label{eq:chiralrot}
\end{align}
As usual, these chiral rotations generate an anomalous coupling to the QCD field strength,
\begin{align}
\mathcal{L}_{aGG}
	=&~ \frac{\alpha_3}{8\pi}\frac{a}{f_a}\,
G_{\mu\nu}\tilde{G}^{\mu\nu}
= \frac{\alpha_3}{8\pi}\frac{a}{n}\frac{3v^2}{v_3^2}\,
G_{\mu\nu}\tilde{G}^{\mu\nu}.
\label{eq:anomaly}
\end{align}
This realizes the Peccei--Quinn mechanism and dynamically removes the effective QCD $\theta$ angle. We therefore identify the PQ scale as
\begin{align}
f_a
=&~ \frac{v_3^2}{3v^2}\,n
\simeq \frac{v_\Sigma}{6}.
\label{eq:faDef}
\end{align}

This relation makes explicit that the axion decay constant is tied to the SU(5)-breaking scale. Since $v_\Sigma$ is the order parameter of SU(5) breaking, it is directly related to the heavy gauge boson mass,
\begin{align}
M_X
= M_{\rm GUT}
=&~ \sqrt{\frac{10\pi}{3}\,\alpha_{\rm GUT}}\,v_\Sigma,
\end{align}
where $M_X$ denotes the mass of the heavy gauge bosons mediating proton decay. Using Eq.~\eqref{eq:faDef}, we finally obtain
\begin{align}
f_a
=&~ \frac{M_{\rm GUT}}{\sqrt{\alpha_{\rm GUT}}}
\frac{1}{2\sqrt{30\pi}}.
\label{eq:faMGUT}
\end{align}
For the relation between the axion mass and the PQ scale, we use the
standard QCD axion relation \cite{Gorghetto:2018ocs},
\begin{align}
m_a
=&~ 5.691(51)\times 10^{-6}\,\mathrm{eV}
\left(\frac{10^{12}\,\mathrm{GeV}}{f_a}\right).
\label{eq:mafa}
\end{align}
Thus, once $M_{\rm GUT}$ and $\alpha_{\rm GUT}$ are fixed, the axion mass range is determined in this framework.

\section{Gauge Coupling Unification}
\label{sec:GCU}

\subsection{Gauge coupling unification with scalar thresholds}

In this section, we briefly review gauge coupling unification (GCU) in the non-supersymmetric SU(5) GUT with a $45$-dimensional Higgs representation \cite{Haba:2024lox}. 
At one loop, the solution of the renormalization group equations (RGEs) relates the gauge couplings at $M_Z$ to those at an energy scale $\mu$ as
\begin{align}
\alpha_i^{-1}(M_Z)
=&~ \alpha_i^{-1}(\mu)
- \frac{B_i}{2\pi}\ln\biggl(\frac{M_Z}{\mu}\biggr).
\end{align}
Here, $M_Z$ denotes the mass of the $Z$ boson, and $B_i$ are the one-loop beta-function coefficients for the three gauge couplings. In terms of
\begin{align}
B_{ij}
=&~ B_i-B_j ,
\end{align}
the one-loop unification condition can be expressed through the combinations $B_{23}/B_{12}$ and $B_{12}$ \cite{GiveonHallSarid1991}.
Using $\alpha_{\rm EM}^{-1}(M_Z)=127.950$, $\sin^2\theta_W(M_Z)=0.23122$, and $\alpha_3(M_Z)=0.1181$, with the SU(5)-normalized hypercharge coupling $\alpha_1=(5/3)\alpha_Y$, successful GCU requires
\begin{align}
\frac{B_{23}}{B_{12}}
=&~ 0.717,
\\
\ln\frac{M_{\rm GUT}}{M_Z}
=&~
\frac{184.95}{B_{12}} .
\end{align}
To realize GCU, we assume a hierarchical mass spectrum for the scalar components contained in $\Phi_{45}$. 
In the numerical analysis, we keep only $S_3$, $S_8$, and one color anti-triplet scalar $S_1$ below the GUT scale. 
The color anti-triplet scalars contained in $\Phi_5$, $\Phi_5^\prime$, and $\Phi_{45}$ can in general mix with one another. 
We assume that the light color anti-triplet mass eigenstate is dominated by the $\Phi_{45}$ component and denote this state by $S_1$:
\begin{align}
	M_{S_3},\,M_{S_8},\,M_{S_1}
	< M_{\rm GUT}.
\end{align}
The remaining scalar multiplets, including the other color anti-triplet mass eigenstates, are taken to have masses of order $M_{\rm GUT}$. 
These intermediate-scale multiplets generate threshold corrections to the one-loop running of the gauge couplings and can lead to successful GCU.
We parametrize the effective one-loop coefficients as
\begin{align}
B_i
=&~
b_i^{\rm SM} + \sum_I b_i^I r_I,
\label{eq:Bi_threshold}
\\
r_I
=&~
\frac{\ln(M_{\rm GUT}/M_I)}{\ln(M_{\rm GUT}/M_Z)} .
\end{align}
where $b_i^{\rm SM}$ are the Standard Model contributions, and $b_i^I$
denote the one-loop contributions of the scalar multiplet $I$.
In the present analysis, we restrict the sum to
\begin{align}
I
=
S_3,\,S_8,\,S_1 .
\end{align}
Here $S_3$ is included only through its threshold contribution to the gauge coupling running; its baryon-number-violating contribution to proton decay is eliminated by imposing the Yukawa condition discussed in Sec.\,\ref{sec:proton decay}. 
Although the adjoint scalar $\Sigma$ is complex in the present PQ realization, the physical non-singlet components of $\Sigma$ are assumed to be degenerate with the heavy states at the GUT scale. 
Their threshold logarithms therefore vanish in the present one-loop GCU analysis. 
Hence, the sum in Eq.~\eqref{eq:Bi_threshold} includes only the intermediate-scale multiplets $S_3$, $S_8$, and $S_1$ arising dominantly from $\Phi_{45}$.
If the adjoint components were split from $M_{\rm GUT}$, their contributions would have to be included with the beta-function coefficients appropriate to complex scalar multiplets.

\begin{table}[t]
\centering
\caption{One-loop contributions of the scalar components in the $\Phi_{45}$ and $\Phi_5^\prime$ representations to the beta-function coefficients. The adjoint components of $\Sigma$ are not included because they are assumed to be degenerate with the heavy GUT scale states in the present analysis.}
\vspace{1mm}
\begin{tabular}{c|ccccccccc}
Fields
& $H^{\prime(5)}$
& $S_1^{\prime(5)*}$
& $\widetilde{S}_1$
& $R_2^*$
& $S_3^*$
& $S_6^*$
& $S_8$
& $H^{(45)}$
& $S_1^{(45)*}$
\\
\hline\hline
$b_1$
& $1/10$
& $1/15$
& $16/15$
& $49/30$
& $1/5$
& $2/15$
& $4/5$
& $1/10$
& $1/15$
\\
\hline
$b_2$
& $1/6$
& $0$
& $0$
& $1/2$
& $2$
& $0$
& $4/3$
& $1/6$
& $0$
\\
\hline
$b_3$
& $0$
& $1/6$
& $1/6$
& $1/3$
& $1/2$
& $5/6$
& $2$
& $0$
& $1/6$
\\
\hline\hline
$B_{12}/r_i$
& $-1/15$
& $1/15$
& $16/15$
& $17/15$
& $-9/5$
& $2/15$
& $-8/15$
& $-1/15$
& $1/15$
\\
\hline
$B_{23}/r_i$
& $1/6$
& $-1/6$
& $-1/6$
& $1/6$
& $3/2$
& $-5/6$
& $-2/3$
& $1/6$
& $-1/6$
\\
\hline
\end{tabular}
\label{tab:betafunctions}
\end{table}

Imposing the unification condition $B_{23}/B_{12}=0.717$, one obtains a linear relation among $r_{S_3}$, $r_{S_8}$, and $r_{S_1}$. 
Substituting this relation into the second unification condition and eliminating
$r_{S_3}$, we obtain
\begin{align}
\ln\frac{M_{\rm GUT}}{M_Z}
=&~ 33.09
- \frac{5}{39}\,\ln\frac{M_{S_8}}{M_Z}
- \frac{1}{78}\,\ln\frac{M_{S_1}}{M_Z}.
\end{align}
Equivalently, this relation can be written as
\begin{align}
\log_{10}\frac{M_{\rm GUT}}{\mathrm{GeV}}
\simeq&~ 16.60
- \frac{5}{39}\,\log_{10}\frac{M_{S_8}}{\mathrm{GeV}}
- \frac{1}{78}\,\log_{10}\frac{M_{S_1}}{\mathrm{GeV}}.
\end{align}

We now determine the unified gauge coupling under the GCU condition. 
Substituting the expression for $\ln(M_{\rm GUT}/M_Z)$ into the RGE for the hypercharge gauge coupling, one obtains
\begin{align}
\alpha_i^{-1}(M_Z)
=&~ \alpha_i^{-1}(M_{\text{GUT}})
- \frac{B_i}{2\pi}\ln\biggl(\frac{M_Z}{M_{\text{GUT}}}\biggr).
\end{align}
For $i=1$, the effective beta function is given by
\begin{align}
B_1
=&~ \frac{351505}{83718}
+ \frac{34340}{41859}\,r_{S_8}
+ \frac{3434}{41859}\,r_{S_1},
\end{align}
where $r_{S_3}$ has again been eliminated using the GCU relation derived above. 
Substituting this into the RGE, we obtain
\begin{align}
\alpha_{\text{GUT}}^{-1}
=&~ \alpha_1^{-1}(M_Z)
- \frac{1}{2\pi}\biggl[
\frac{175518783}{1040000}
- \frac{115}{78}\,
\ln\frac{M_{S_8}}{M_Z}
- \frac{23}{156}\,
\ln\frac{M_{S_1}}{M_Z}
\biggr]
\nonumber\\
\simeq&~ 30.99
+ 0.2347\,\ln\frac{M_{S_8}}{\mathrm{GeV}}
+ 0.0235\,\ln\frac{M_{S_1}}{\mathrm{GeV}}
\nonumber\\
\simeq&~ 30.99
+ 0.5404\,\log_{10}\frac{M_{S_8}}{\mathrm{GeV}}
+ 0.0541\,\log_{10}\frac{M_{S_1}}{\mathrm{GeV}}.
\end{align}
Together with the expression for $M_{\rm GUT}$, this result fixes $\alpha_{\rm GUT}$ in terms of the light-scalar threshold masses.
The resulting values of $M_{\rm GUT}$ and $\alpha_{\rm GUT}$ will be used below to evaluate the proton decay rates and to determine the axion mass through the axion decay constant.

Proton decay provides a direct probe of the GUT scale and thereby severely restricts the parameter space. 
After eliminating $r_{S_3}$, the GCU conditions fix $M_{\rm GUT}$ and $\alpha_{\rm GUT}$ as functions of $M_{S_8}$ and $M_{S_1}$. 
Consequently, the experimental limits on proton decay can be mapped onto the $(M_{S_8}, M_{S_1})$ plane and impose an upper bound on $M_{S_8}$ for each value of $M_{S_1}$.

\subsection{Gauge-boson-mediated proton decay constraints}

We first consider the antineutrino mode,
\begin{align}
\Gamma(p\to K^+\bar\nu)
=&~ \frac{\pi m_p}{2}\,
\frac{\alpha_{\rm GUT}^{2}}{M_{\rm GUT}^{4}}\,
\biggl(1 - \frac{m_{K^+}^2}{m_p^2}\biggr)^{2}
C_K .
\end{align}
The factor $C_K$ collects the hadronic matrix elements and renormalization effects:
\begin{align}
C_K
=&~ A_{RG}^2
\biggl[
\bigl|(V_{\rm CKM})_{11}\braket{K^+|(us)_R d_L|p}\bigr|^2
+
\bigl|(V_{\rm CKM})_{12}\braket{K^+|(ud)_R s_L|p}\bigr|^2
\biggr]
\simeq 0.01 .
\end{align}
A key feature of this mode is its independence from the charged-lepton flavor rotation matrix. 
The $p\to K^+\bar\nu$ mode therefore provides the most robust proton decay constraint available in SU(5) without additional assumptions on the charged-lepton flavor structure. 
The dependence on $M_{S_8}$ enters only through $M_{\rm GUT}$ and $\alpha_{\rm GUT}$, which are fixed by the GCU conditions after eliminating $r_{S_3}$.

We next consider the charged-lepton mode,
\begin{align}
\Gamma (p \to e^+ \pi^0)
=&~ \frac{\pi m_p}{2}\left(1-\frac{m_\pi^2}{m_p^2}\right)^2
\frac{\alpha_{\rm GUT}^2}{M_{\rm GUT}^4}
A_{RG}^2
V_e^2
\bigl|\braket{\pi^0|(ud)_R u_L|p}\bigr|^2 ,
\end{align}
where
\begin{align}
V_e
=&~ \sqrt{\bigl|
V_2^{11} + V_{\rm CKM}^{11}(V_2V_{\rm CKM})^{11}
\bigr|^2
+ \bigl|V_3^{11}\bigr|^2}.
\end{align}
Unlike the antineutrino mode, this mode depends on the charged-lepton flavor structure through $V_e$. 
This mode has the most stringent experimental lower bound among the modes listed in Table~\ref{Tab:ProtonLifeTime}.
\begin{table}[h]
  \caption{Experimental constraints on the lifetimes of various proton decay modes by Super-Kamiokande \cite{Super-Kamiokande:2020wjk,Super-Kamiokande:2013rwg,Super-Kamiokande:2022egr,Super-Kamiokande:2014otb,Super-Kamiokande:2005lev}.}
  \begin{tabular}{|c|c|c|c|c|c|c|} 
   \hline 
   Decay mode & $p\rightarrow\mu^+K^0$ & $p\rightarrow\mu^+\pi^0$ & $p\rightarrow e^+K^0$ & $p\rightarrow e^+\pi^0$ & $p\rightarrow K^+\overline{\nu}$ & $p\rightarrow \pi^+\overline{\nu}$  \\\hline
	90\% C.L. (years) & $3.6\times10^{33}$ & $1.6\times10^{34}$ & $1.0\times10^{33}$ & $2.4\times10^{34}$ & $5.9\times10^{33}$ &$3.9\times10^{32}$ 
	\\\hline
  \end{tabular}
  \label{Tab:ProtonLifeTime}
\end{table}

Since $V_e$ cannot be fixed within SU(5) alone, the lifetime of this mode is not uniquely predicted unless one specifies an additional flavor ansatz. 
To illustrate the impact of this flavor uncertainty, we adopt the benchmark values
\begin{align}
V_e
=&~ 0.1,\ 1.0,\ 2.2 .
\end{align}
The value $V_e=2.2$ represents a case in which the relevant flavor factors are of order unity.

\begin{figure}[H]
\centering
\includegraphics[width=80mm]{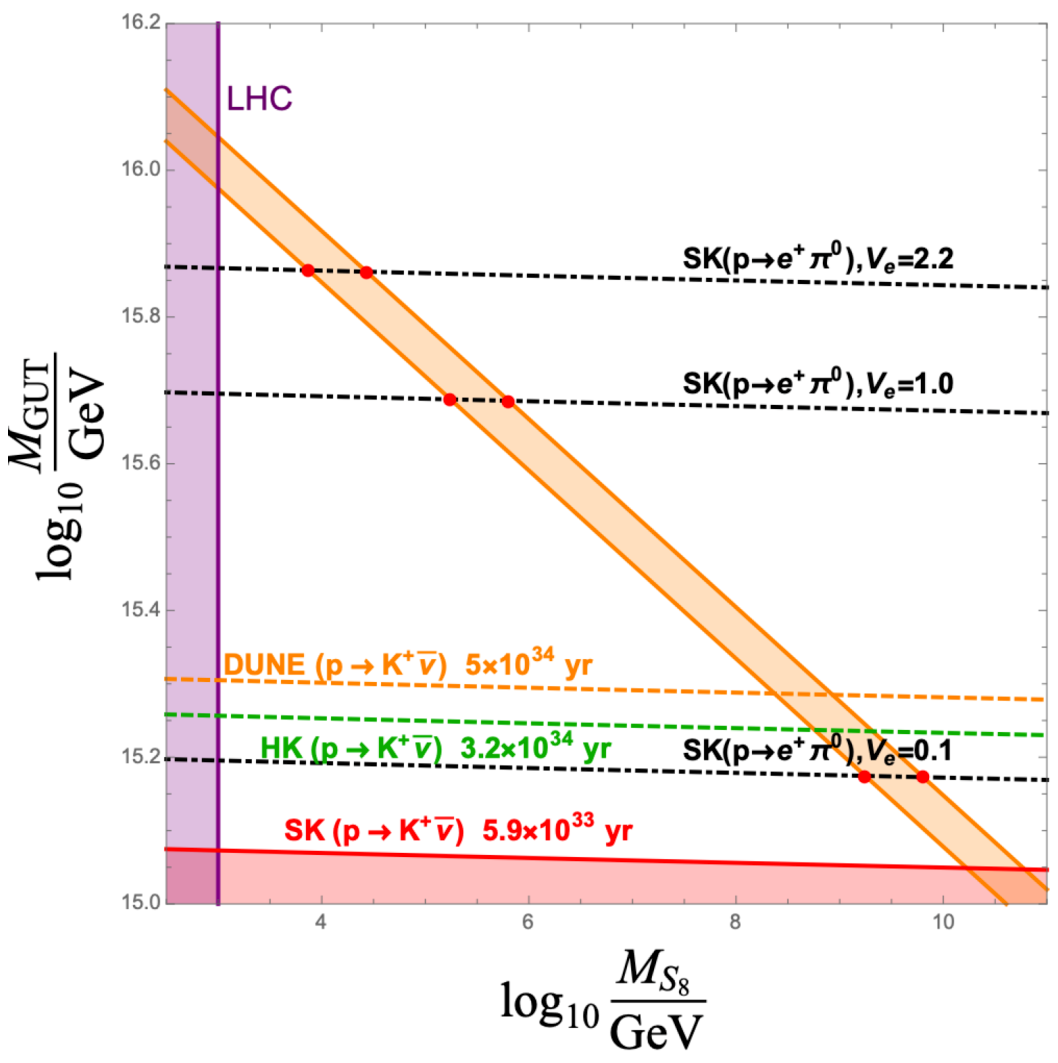}
\vspace{-4mm}
\caption{
Parameter region in the $M_{S_8}$--$M_{\rm GUT}$ plane. 
The orange band represents the region where GCU is realized, with its boundaries corresponding to $M_{S_1}=3.4\times10^{13}\,{\rm GeV}$ and $M_{S_1}=M_{\rm GUT}$. 
The purple shaded region is excluded by the LHC bound. 
The colored dashed lines denote the current bound and the projected sensitivity from $p\to K^+\bar{\nu}$. 
The black dashed lines show the current bounds from $p\to e^+\pi^0$ for $V_e=0.1$, $1.0$, and $2.2$.
}
\label{fig:82region}
\end{figure}

Fig.\,\ref{fig:82region} shows the predictions for $p \to e^+ \pi^0$, together with the Super-Kamiokande bound,
$\tau/B(p\to e^+ \pi^0)>2.4 \times 10^{34}$ years~\cite{Super-Kamiokande:2020wjk},
and the projected Hyper-Kamiokande sensitivity,
$\tau/B(p\to e^+ \pi^0)>8 \times 10^{34}$ years~\cite{Yokoyama:2017mnt}.
Even when $M_{S_1}$ is varied over the range
\begin{align}
3.4\times10^{13}\,{\rm GeV}
\leq&~ M_{S_1}
\leq M_{\rm GUT},
\end{align}
the resulting bound on $M_{S_8}$ changes only mildly. 
Thus, once a definite flavor structure is specified, the charged-lepton mode can provide an additional and potentially stronger constraint on the allowed parameter space.

The two classes of proton decay modes therefore provide complementary constraints on the model. 
The antineutrino modes are independent of the charged-lepton flavor rotation matrix and yield the most conservative bounds that can be derived within SU(5) alone. 
By contrast, the charged-lepton modes depend on the flavor parameter $V_e$; thus, their constraints require an additional assumption about the charged-lepton flavor structure. 
For sufficiently large $V_e$, these modes can impose stronger constraints on $M_{\mathrm{GUT}}$, and hence on $M_{S_8}$, than the antineutrino modes.

\section{Proton Decay Mediated by the 45-dimensional Higgs\label{sec:proton decay}}

In the gauge coupling unification analysis, the scalar components of $\Phi_{45}$ can appear as intermediate-scale thresholds. 
In particular, when $S_1$, $S_3$, and $S_8$ are significantly lighter than the GUT scale, they can modify the running of the gauge couplings. 

The $S_3$ exchange generates a particularly dangerous contribution to proton decay because the corresponding amplitudes involve the two Yukawa couplings $Y_3^{QQ}$ and $Y_3^{QL}$. 
The current limit on $p\to e^+\pi^0$ implies the following constraint on the $S_3$ Yukawa couplings:
\begin{align}
\bigl|
(Y_3^{QQ})_{12}
(Y_3^{QL})_{11}
\bigr|
&<~
10^{-17}\,s_H
\label{eq:Y3QQY3QL}
\end{align}
for $m_{S_3}=5\times10^7\,{\rm GeV}$ \cite{Haba:2024lox}.
Here $s_H\equiv \sin\theta_H$, where $\theta_H$ denotes the mixing angle between the electroweak-doublet components in $\Phi_5$ and $\Phi_{45}$.
Thus, $s_H$ measures the fraction of the $45$-dimensional Higgs doublet contained in the SM-like Higgs doublet.
Using the matching conditions between the GUT and SM Yukawa couplings, this constraint can be translated into
\begin{align}
\bigl|
(Y_{45}^U)_{12}
\bigr|
<&~
10^{-12}s_H .
\end{align}
To suppress the $S_3$-mediated proton decay contribution, we impose
\begin{align}
Y_{45}^U
=&~0
\label{eq:Y45U_zero}
\end{align}
at the GUT scale~\cite{Haba:2024lox,Goto:2023qch}. 
Even with this assumption, realistic fermion masses and mixings can be reproduced in the extended GUT framework.

In general, both $S_1$ and $S_3$ can mediate baryon-number-violating processes~\cite{Dorsner:2009cu,Dorsner:2012nq}. 
In the present setup, however, the baryon-number-violating contribution of $S_3$ is absent because the relevant Yukawa coupling is set to zero by Eq.~\eqref{eq:Y45U_zero}. 
Therefore, throughout the parameter scan, $S_3$ is retained only as an intermediate-scale threshold entering the gauge coupling unification conditions. The bound in Eq.~\eqref{eq:Y3QQY3QL} is quoted only to illustrate the severity of the generic $S_3$-mediated proton decay constraint, and is not imposed as an additional constraint in the scan.

We therefore focus on the proton decay constraints induced by the exchange of the $S_1$ component. 
The scalar $S_8$ affects the present analysis only through the gauge coupling threshold correction. 
In the proton decay calculation, we assume that the mixing between the $S_1$ component in $\Phi_{45}$ and the color-triplet component in $\Phi_5$ is sufficiently small and can be neglected. 
The lower bound on $M_{S_1}$ obtained from $S_1$-mediated proton decay directly affects the present analysis because the GCU condition correlates $M_{S_1}$, $M_{S_8}$, and $M_{\rm GUT}$.
Consequently, this bound removes part of the otherwise allowed region in the $M_{S_8}$--$M_{\rm GUT}$ plane. 
In this section, we summarize the $S_1$-mediated proton decay constraint and impose it in the GCU parameter-space analysis below.

The size and flavor structure of the $S_1$ couplings are fixed by the Georgi--Jarlskog structure. 
In the flavor basis, the 45-dimensional Higgs field is assumed to contribute only to the second-generation entry. 
This entry generates the parameter $C$ in the down-quark and charged-lepton mass matrices, leading to the GUT scale relations
\begin{align}
m_s \simeq |C|,\qquad
m_\mu \simeq 3|C|.
\end{align}
Thus, up to coefficients of order unity and representation-dependent
normalization factors, the relevant entry of $Y_{45}^D$ is estimated as
\begin{align}
(Y_{45}^D)_{22}
\sim&~
\frac{v}{v_{45}}\lambda^2 .
\end{align}
Rotating this coupling to the fermion mass basis gives the $S_1$ couplings
\begin{align}
\bigl(Y_1^{DU}\bigr)^T
\sim&~
\frac{v}{v_{45}}
\begin{pmatrix}
\lambda^5 & \lambda^4 & 0 \\
\lambda^3 & \lambda^2 & 0\\
\lambda^5 & \lambda^4 & 0
\end{pmatrix}, \,\,\,
Y_1^{QL,\ell}
\sim 
\frac{v}{v_{45}}
\begin{pmatrix}
\lambda^5 & \lambda^4 & 0\\
\lambda^3 & \lambda^2 & 0\\
\lambda^5 & \lambda^4 & 0
\end{pmatrix}, \,\,\,
Y_1^{QL,\nu}
\sim 
\frac{v}{v_{45}}
\begin{pmatrix}
\lambda^4 & \lambda^3 & 0\\
\lambda^3 & \lambda^2 & 0\\
0 & 0 & 0
\end{pmatrix}.
\end{align}

These estimates determine the flavor factors entering the $S_1$-mediated proton decay amplitudes. 
The derivation of the texture estimates and the corresponding flavor rotations is given in Appendix \ref{app:S1_proton_decay}.

The $S_1$ exchange generates baryon-number-violating four-fermion operators whose coefficients are proportional to the product of the two $S_1$ Yukawa couplings.
After rotating to the fermion mass basis, we define
\begin{align}
\delta_{ijkl}^{L,\ell}
=&~
\bigl(Y_1^{DU}\bigr)_{ij}^{*}
\bigl(Y_1^{QL,\ell}\bigr)_{kl},
\label{eq:deltaL_lepton_def}
\\
\delta_{ijkl}^{L,\nu}
=&~
\bigl(Y_1^{DU}\bigr)_{ij}^{*}
\bigl(Y_1^{QL,\nu}\bigr)_{kl}.
\label{eq:deltaL_neutrino_def}
\end{align}
The two matrices $Y_1^{QL,\ell}$ and $Y_1^{QL,\nu}$ denote the charged-lepton and neutrino components of the same $SU(2)_L$ invariant coupling after rotating to the fermion mass basis. 
These matrices are not independent Yukawa matrices.

The quantities $\delta_{ijkl}^{L,\ell}$ and $\delta_{ijkl}^{L,\nu}$ encode the flavor suppression factors of the proton decay amplitudes. 
Since the $S_1$ exchange generates amplitudes proportional to $\delta^L/M_{S_1}^2$, the corresponding partial widths scale as $|\delta^L|^2/M_{S_1}^4$. 
For the Georgi--Jarlskog texture considered here, the strongest constraint is obtained from
\begin{align}
\Gamma(p\to K^+\bar{\nu})
=&~
\frac{1}{64\pi}
\left(1-\frac{m_K^2}{m_p^2}\right)^2
\frac{m_p}{f^2}
\alpha_H^2 A_{RL}^2
\frac{1}{M_{S_1}^4}
\sum_{i=1}^{3}
\left|
\delta_{211i}^{L,\nu}
\frac{2D}{3}
+
\delta_{112i}^{L,\nu}
\left(1+\frac{D}{3}+F\right)
\right|^2 .
\end{align}
In this texture, the neutrino-flavor sum is dominated by the $i=2$ contribution. 
Here we follow the notation and input parameters summarized in Appendix~\ref{app:S1_proton_decay}. 
The full set of partial widths used in the numerical analysis is also given there.

The remaining charged-lepton and antineutrino modes are evaluated in Appendix \ref{app:S1_proton_decay}. 
Their flavor suppressions, experimental lifetime limits, and the resulting lower bounds on $M_{S_1}$ are summarized in Table~\ref{tab:proton decay}.
\begin{table}[H]
  \caption{Lower limits of the mass of the $S_1$ boson obtained from the 90\% CL bounds on proton partial lifetimes \cite{Super-Kamiokande:2020wjk,Super-Kamiokande:2013rwg,Super-Kamiokande:2022egr,Super-Kamiokande:2014otb,Super-Kamiokande:2005lev}.}
  \vspace{1mm}
  \begin{tabular}{|c|c|c|c|c|c|c|}
\hline
   Decay mode & $p\rightarrow\mu^+K^0$ & $p\rightarrow\mu^+\pi^0$ & $p\rightarrow e^+K^0$ & $p\rightarrow e^+\pi^0$ & $p\rightarrow K^+\overline{\nu}$ & $p\rightarrow \pi^+\overline{\nu}$\\ \hline
   Partial width & $\propto(\lambda^8)^2$ & $\propto(\lambda^9)^2$ & $\propto(\lambda^9)^2$ & $\propto(\lambda^{10})^2$ & $\propto(\lambda^7)^2$ & $\propto(\lambda^8)^2$ \\ \hline
	90\% CL bound [years]& $3.6\times10^{33}$ & $1.6\times10^{34}$ & $1.0\times10^{33}$ & $2.4\times10^{34}$ & $5.9\times10^{33}$ & $3.9\times10^{32}$  \\ \hline
	Lower limit [GeV]& $1.1\times10^{13}$ & $9.5\times10^{12}$ & $3.7\times10^{12}$ & $4.9\times10^{12}$ & $3.4\times10^{13}$ & $9.5\times10^{12}$ \\ \hline 
    \end{tabular}
  \label{tab:proton decay}
\end{table}
\noindent
As shown in Table~\ref{tab:proton decay}, the strongest constraint comes from the $p\to K^+\bar{\nu}$ mode. 
In the Georgi--Jarlskog benchmark with $v/v_{45}=\sqrt{2}$, this constraint implies
\begin{align}
M_{S_1}
\geq&~3.4\times10^{13}\,{\rm GeV}.
\end{align}
Since $S_1$ is treated as an intermediate-scale threshold in the GCU analysis, we also require $M_{S_1}\leq M_{\rm GUT}$. 
We therefore impose
\begin{align}
3.4\times10^{13}\,{\rm GeV}
\leq&~M_{S_1}
\leq M_{\rm GUT}.
\end{align}

\begin{figure}[H]
    \includegraphics[width=56mm]{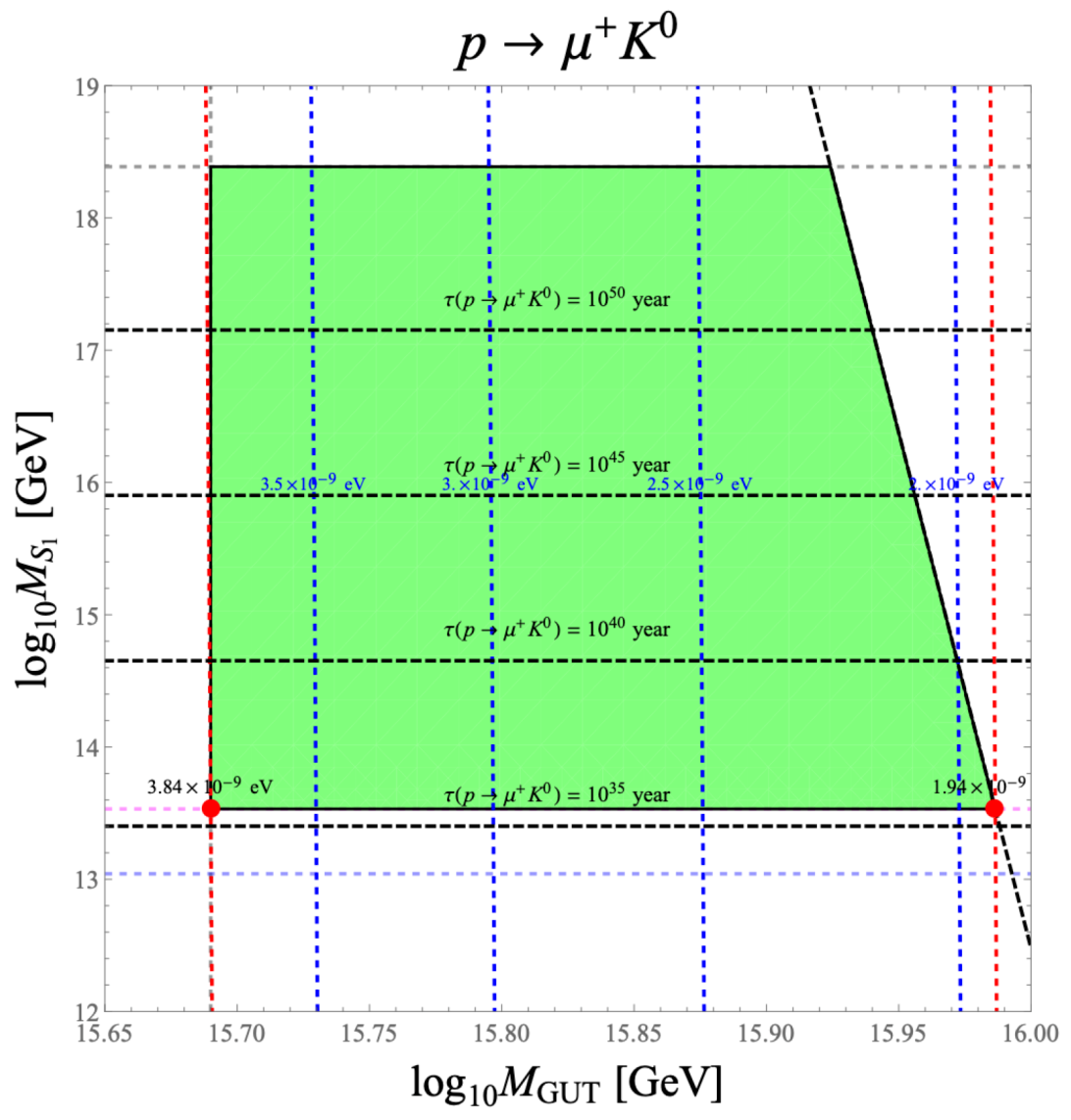}
    \includegraphics[width=56mm]{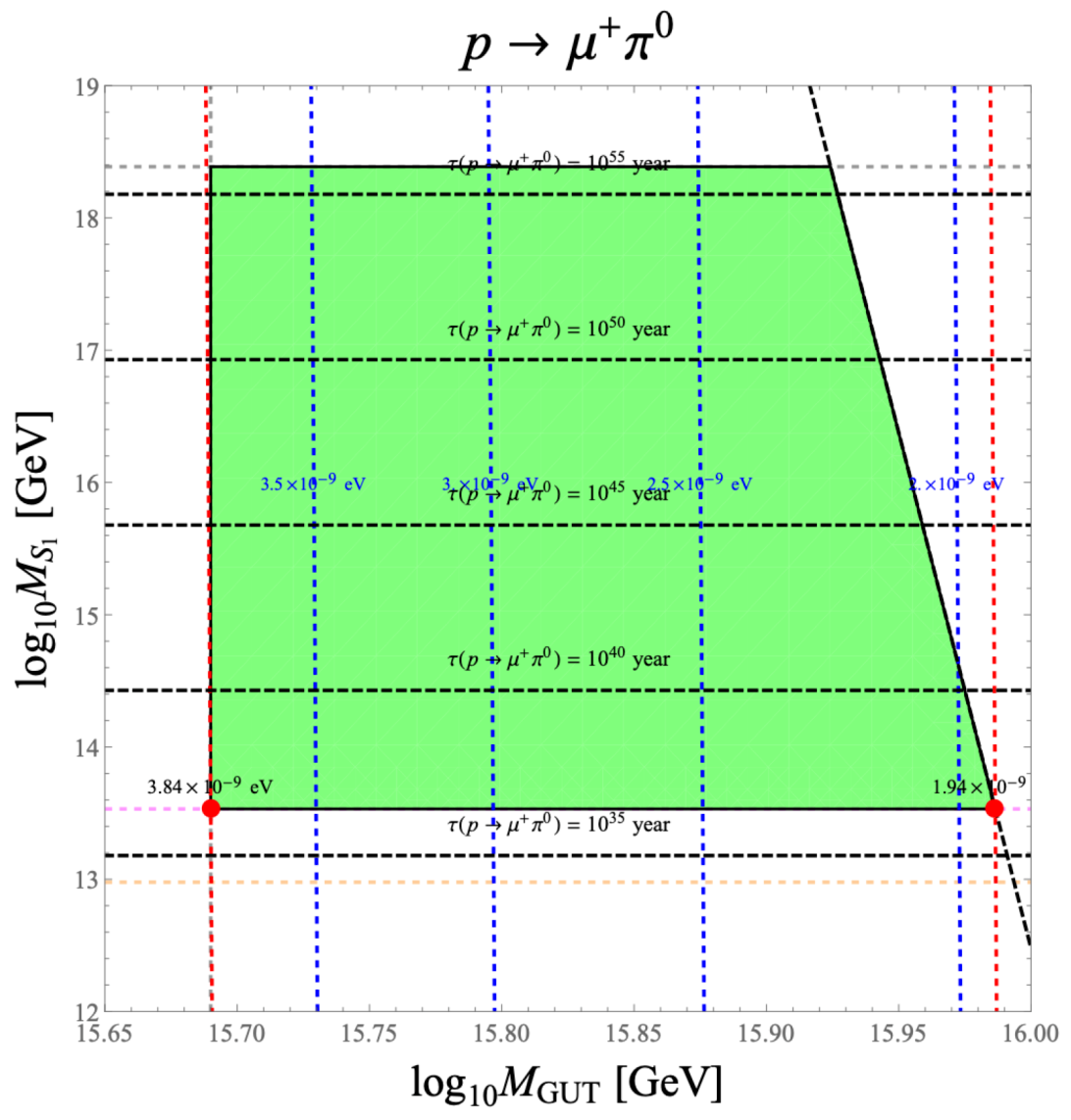}
    \includegraphics[width=56mm]{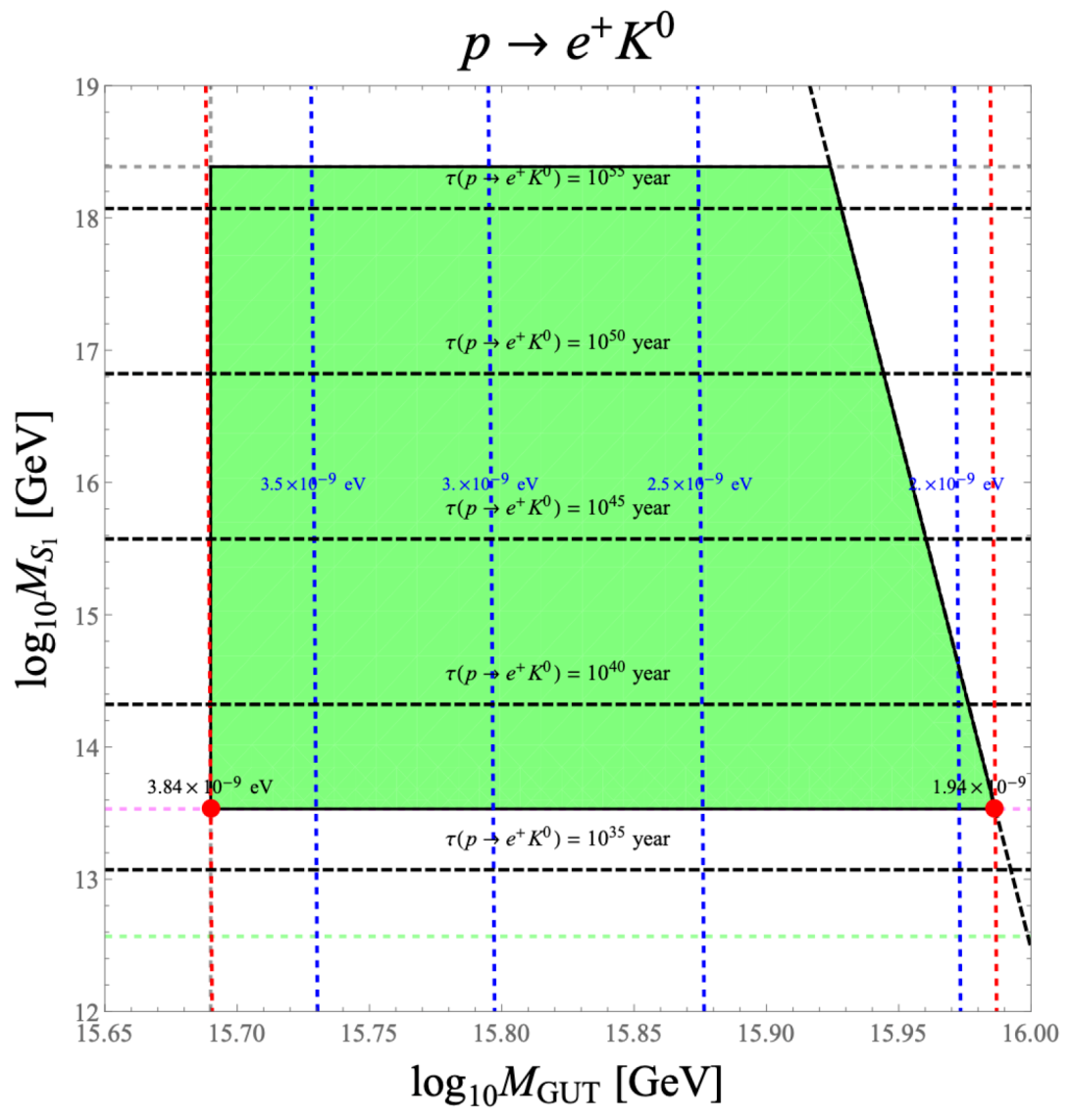}
    \includegraphics[width=56mm]{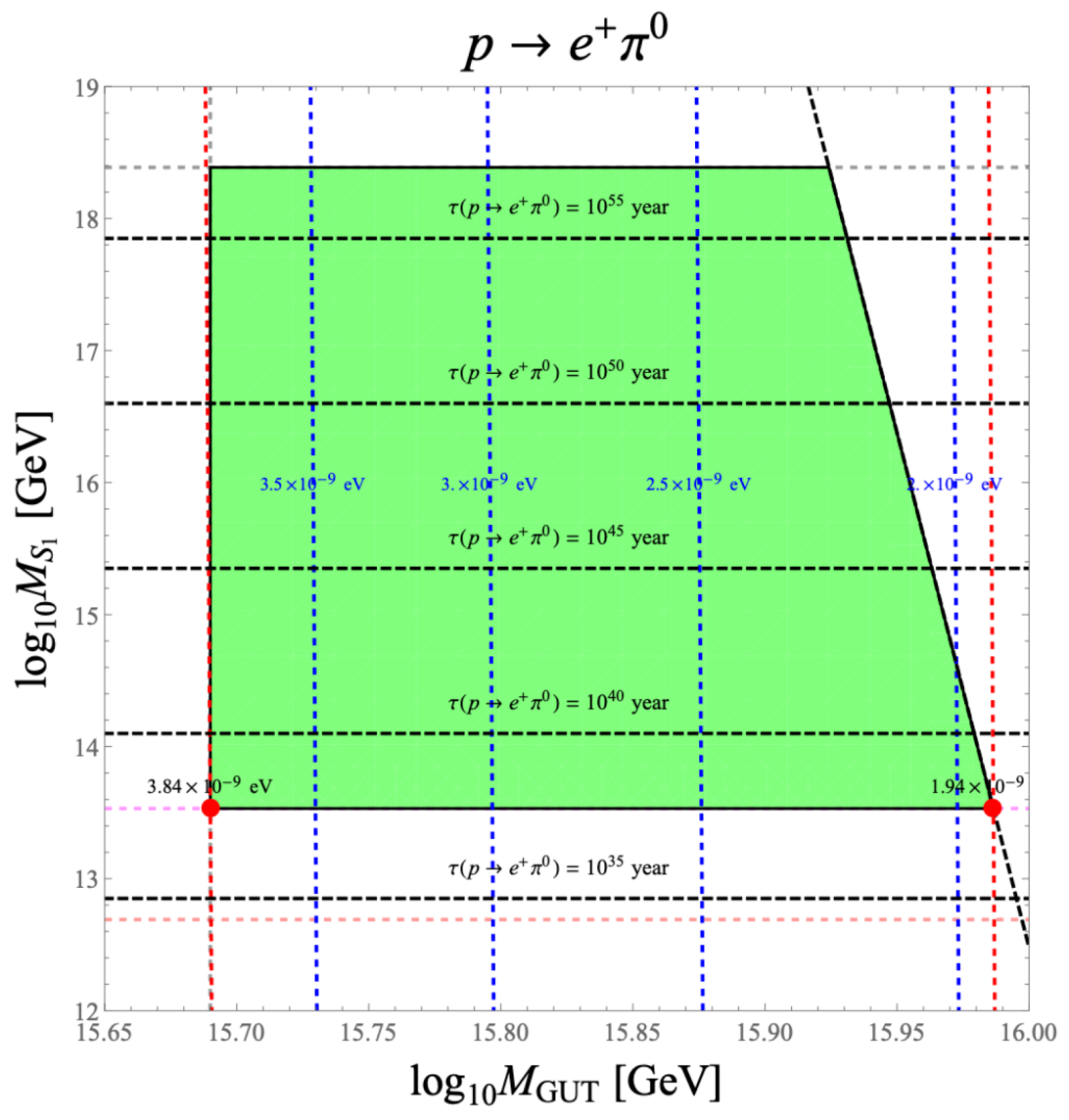}
    \includegraphics[width=56mm]{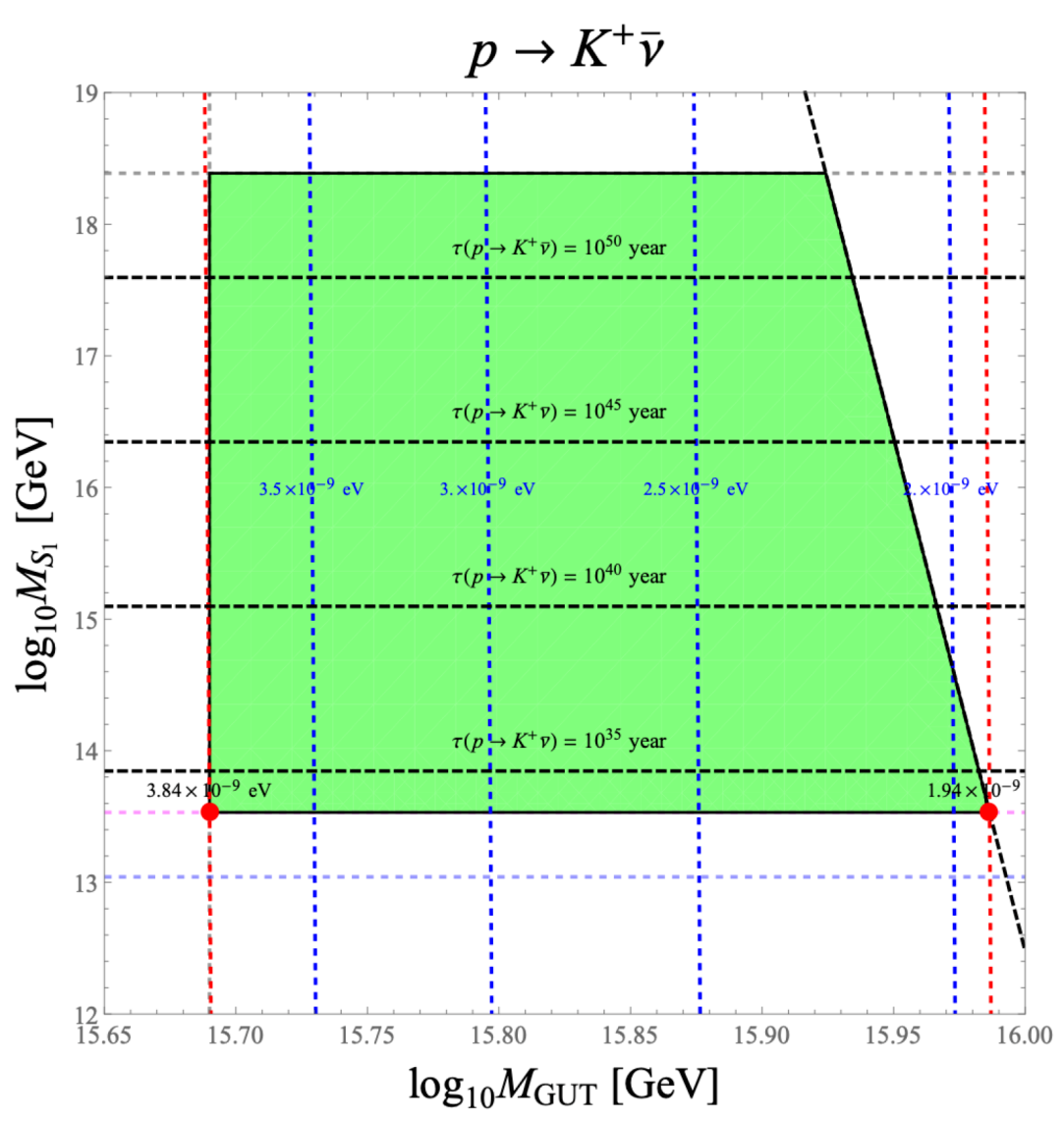}
    \includegraphics[width=56mm]{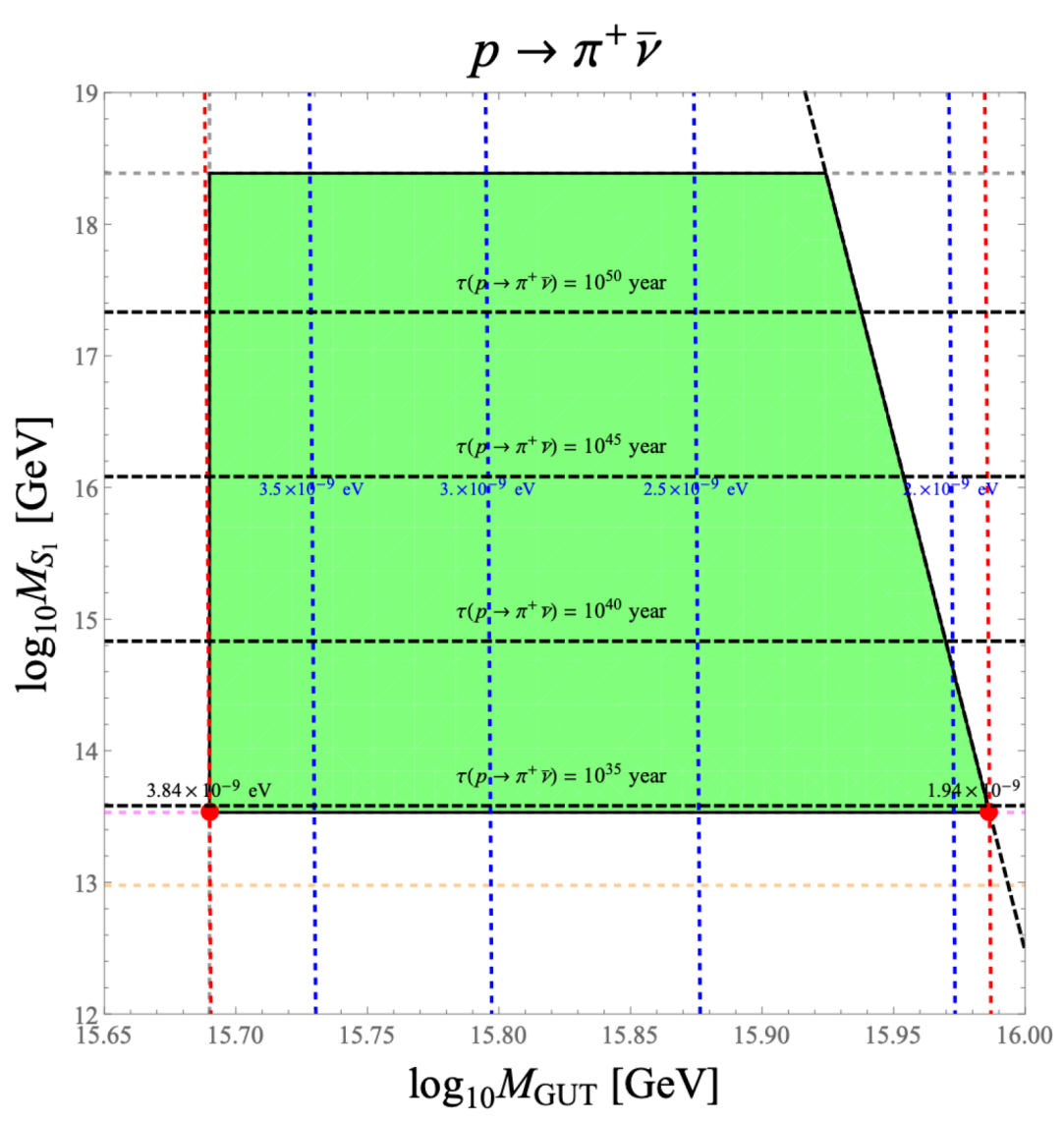}
    \vspace{-7mm}
    \caption{
Allowed regions in the $M_{\mathrm{GUT}}$--$M_{S_1}$ plane for the six proton decay modes considered in this analysis.
The decay mode for each panel is indicated above the corresponding panel.
The green shaded regions are allowed by gauge coupling unification and the current experimental constraints.
The horizontal black dashed lines show the projected lower bounds on $M_{S_1}$ inferred from future sensitivities to the corresponding partial proton lifetimes $\tau/B$.
The vertical blue dashed lines correspond to fixed values of the axion mass $m_a$.
The red dashed lines and red points indicate the endpoints of the allowed axion-mass interval.
The black slanted boundary corresponds to $M_{S_8}=3 \,\mathrm{TeV}$; the region requiring $M_{S_8}<3 \,\mathrm{TeV}$ is excluded by the lower bound on $M_{S_8}$~\cite{Hayreter:2017wra}.
}
\label{fig:allowed_M31_MGUT}
\end{figure}

Fig.\,\ref{fig:allowed_M31_MGUT} shows the allowed region in the $M_{\rm GUT}$--$M_{S_1}$ plane in the scenario with three light scalar components. 
The colored region satisfies the GCU condition and the proton decay constraints. 
The horizontal and vertical axes denote $M_{\rm GUT}$ and $M_{S_1}$, respectively, and both are shown on logarithmic scales. 
The diagonal lines with negative slope indicate contours of fixed $M_{S_8}$.

The left boundary of the allowed region is set by gauge-boson-mediated proton decay. Among the current limits, the $p\rightarrow e^+\pi^0$ mode provides the strongest constraint on $M_{\rm GUT}$. 
Since this mode depends on the charged-lepton flavor rotation matrix $V_e$, we show the allowed regions for the representative values $V_e=0.1$, $1.0$, and $2.2$. 
By contrast, the antineutrino modes are summed over the final-state neutrino flavor and do not depend on this particular SU(5) mixing matrix.

For the three representative values of $V_e$, the lower edge of the allowed region corresponds to the following values of the unified gauge coupling and the GUT scale:
\begin{align}
	V_e=0.1:~~
	\alpha_{\text{GUT}}^{-1}
	\simeq&~ 36.9806,
	&
	\log_{10}\left(\frac{M_{\text{GUT}}}{\text{GeV}}\right)
	\simeq&~ 15.1769,
	&
	M_{\text{GUT}}
	\simeq&~ 1.50\times 10^{15}~\text{GeV},
	\\[1.5mm]
	V_e=1.0:~~
	\alpha_{\text{GUT}}^{-1}
	\simeq&~ 34.8179,
	&
	\log_{10}\left(\frac{M_{\text{GUT}}}{\text{GeV}}\right)
	\simeq&~ 15.6900,
	&
	M_{\text{GUT}}
	\simeq&~ 4.90\times 10^{15}~\text{GeV},
	\\[1.5mm]
	V_e=2.2:~~
	\alpha_{\text{GUT}}^{-1}
	\simeq&~ 34.0765,
	&
	\log_{10}\left(\frac{M_{\text{GUT}}}{\text{GeV}}\right)
	\simeq&~ 15.8659,
	&
	M_{\text{GUT}}
	\simeq&~ 7.34\times 10^{15}~\text{GeV}.
\end{align}
Thus, a larger value of $V_e$ strengthens the $p\rightarrow e^+\pi^0$ constraint and shifts the allowed region toward larger $M_{\text{GUT}}$.

The lower boundary in the $M_{S_1}$ direction is determined by scalar-mediated proton decay. 
Using the experimental limits summarized in Table~\ref{tab:proton decay}, the strongest present constraint comes from the $p\rightarrow K^+\overline{\nu}$ mode and implies $M_{S_1}\geq 3.4\times10^{13}\,{\rm GeV}$. 
This lower bound is imposed in the GCU analysis. 
The region above $M_{S_1}=2.4\times10^{18}\,{\rm GeV}$ is excluded because the scalar mass would exceed the Planck scale.

The right boundary is determined by the lower bound on $M_{S_8}$.
We impose $M_{S_8}>3\,{\rm TeV}$, as motivated by LHC searches \cite{Hayreter:2017wra}. 
Since the GCU condition relates $M_{S_8}$ to $M_{\rm GUT}$ and $M_{S_1}$, this bound excludes the region in which the value of $M_{S_8}$ required by unification becomes smaller than $3\,{\rm TeV}$. 
The allowed range of $M_{S_8}$ therefore depends indirectly on $V_e$, because the charged-lepton proton decay constraint fixes the lower edge of the allowed $M_{\rm GUT}$ region for each value of $V_e$.
After imposing the GCU condition, the proton decay constraints, and the LHC bound on $M_{S_8}$, we obtain
\begin{align}
	V_e=0.1:~~
	3 ~ \text{TeV}
	\leq &~ M_{S_8} \leq 2.1\times 10^{9}~\text{GeV},
	\\
	V_e=1.0:~~
	3 ~ \text{TeV}
	\leq &~ M_{S_8} \leq 2.1\times 10^{5}~\text{GeV},
	\\
	V_e=2.2:~~
	3 ~ \text{TeV}
	\leq &~ M_{S_8} \leq 8.7\times 10^{3}~\text{GeV}.
\end{align}
Thus, a larger value of $V_e$ raises the lower bound on $M_{\rm GUT}$ through the $p\rightarrow e^+\pi^0$ constraint and reduces the allowed range of $M_{S_8}$.

The lower-left corner of the allowed region also fixes the largest allowed axion mass. 
At this point, $M_{\rm GUT}$ reaches its smallest allowed value, and the GUT--PQ relation gives the smallest allowed $f_a$. 
Since the axion mass decreases as $f_a$ increases, the largest axion mass is obtained at the smallest value of $M_{\rm GUT}$ allowed by the proton decay constraint. 
A larger value of $V_e$ strengthens this lower bound on $M_{\rm GUT}$, thereby raising the corresponding lower bound on $f_a$ and reducing the maximum allowed axion mass.

The horizontal dashed lines show the impact of future proton decay sensitivities in the scalar-mediated modes. 
In the Georgi--Jarlskog setup, improved limits on these modes raise the lower bound on $M_{S_1}$ and exclude the parameter space from below. 
Improved sensitivity to $p\rightarrow e^+\pi^0$ raises the lower bound on $M_{\rm GUT}$ and excludes the parameter space from the left. 
Future proton decay searches can therefore reshape the allowed $M_{\rm GUT}$--$M_{S_1}$ region and, through the GUT--PQ relation, the predicted axion mass window.

\section{Axion phenomenology}
\label{sec:axion}

In this section, we evaluate the axion observables in the allowed region selected by the proton decay and gauge coupling unification constraints. 
Since $f_a$ is related to $M_{\rm GUT}$ and $\alpha_{\rm GUT}$, these constraints restrict the allowed ranges of $m_a$, $g_{a\gamma\gamma}$, and the axion-induced neutron electric dipole moment (EDM).

The cosmological history of PQ breaking determines whether the domain-wall issue associated with the PQ symmetry becomes relevant. 
In the present work, we restrict our analysis to histories in which the PQ symmetry is broken before or during inflation and is not restored after inflation. 
Then inflation selects a single PQ vacuum, and no stable string--wall network remains in the observable Universe.

With the standard integer normalization of the PQ charges, the color anomaly coefficient is
\begin{align}
C_{ag}
=&~
2\sum_f X_f T(R_f)
\nonumber\\
=&~
3(\alpha+3\beta)
\nonumber\\
=&
-6X_\Sigma .
\end{align}
Here $X_\Sigma=-(\alpha+3\beta)/2$ is the PQ charge of the adjoint field.
Normalizing the dominant PQ breaking field as $|X_\Sigma|=1$, we obtain
\begin{align}
N_{\rm DW}
=&~
|C_{ag}|
=
6 .
\end{align}
Thus, a post-inflationary PQ breaking history would lead to the usual DFSZ domain-wall problem, whereas the pre-inflationary history assumed here avoids the formation of a stable string--wall network in the observable Universe.

For the large decay constant predicted in the present model, $f_a \sim 10^{14}$--$10^{15}\,{\rm GeV}$, an initial misalignment angle of order unity would generically overproduce axion dark matter in the standard misalignment scenario. 
We therefore assume that the initial misalignment angle is sufficiently small. We also assume that inflationary isocurvature constraints are avoided, for example by a sufficiently low inflationary scale.
Under these assumptions, the coherent axion field can constitute all or part of the local dark-matter background relevant for oscillating-EDM searches. 
If axions account for only a fraction of the local dark-matter density, the oscillating EDM amplitude should be rescaled by
\begin{align}
\left(
\frac{\rho_a}{\rho_{\rm DM}}
\right)^{1/2}.
\end{align}

The allowed range of $f_a$ is restricted by the values of $M_{\rm GUT}$ and $\alpha_{\rm GUT}$ obtained from the gauge coupling unification analysis, together with the proton decay constraints. 
Using Eq.~\eqref{eq:faMGUT}, $f_a$ is fixed once $M_{\rm GUT}$ and $\alpha_{\rm GUT}$ are specified. 
The axion mass is then obtained from Eq.~\eqref{eq:mafa}. 
After imposing the GCU condition, it can be written in terms of $M_{\rm GUT}$, $M_{S_8}$, and $M_{S_1}$ as
\begin{align}
m_a
\simeq&~
1.105\times 10^{-4}\,{\rm eV}\,
\frac{1}{
\sqrt{\displaystyle
30.99
+ 0.5404\,\log_{10}\frac{M_{S_8}}{{\rm GeV}}
+ 0.0541\,\log_{10}\frac{M_{S_1}}{{\rm GeV}}
}
}
\left(
\frac{10^{12}\,{\rm GeV}}{M_{\rm GUT}}
\right).
\end{align}

\begin{figure}[t]
    \includegraphics[width=78mm]{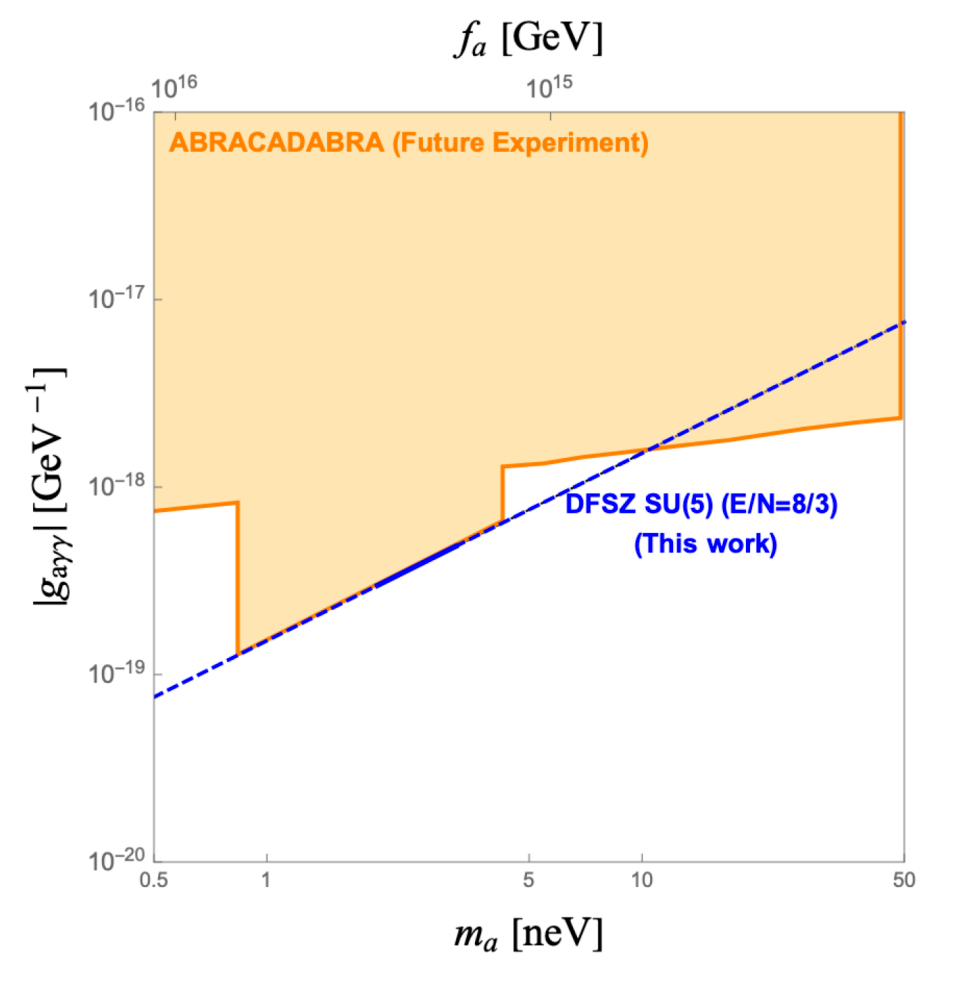}~~
    \includegraphics[width=78mm]{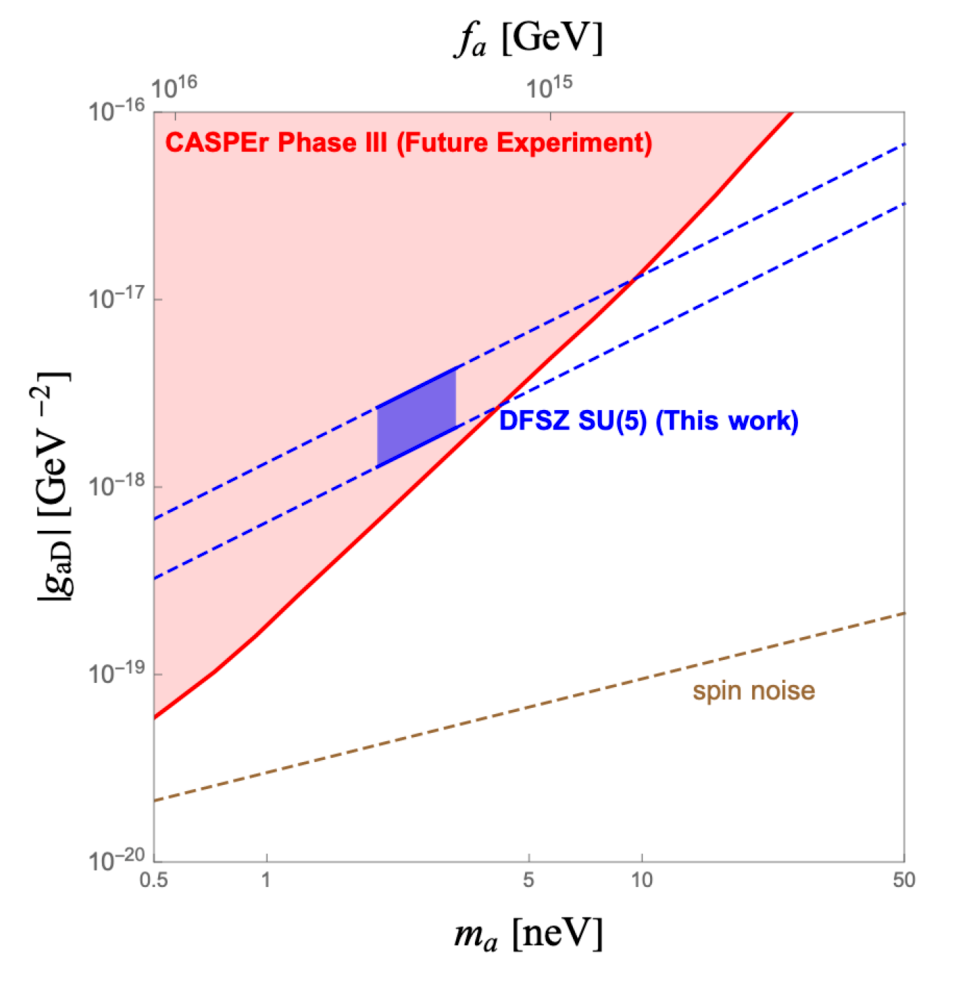}
    \vspace{-6mm}
  \caption{Axion predictions for the representative case $V_e=1.0$. 
Left panel: prediction in the $m_a$--$g_{a\gamma\gamma}$ plane.
The blue solid line shows the prediction of the present model with $E/N=8/3$. 
The orange shaded region indicates the projected ABRACADABRA Phase-III broadband sensitivity, while the orange boundary line denotes the corresponding resonant reach under the benchmark assumptions described in the text. \\
Right panel: prediction in the $m_a$--$g_{aD}$ plane.
The blue band shows the prediction of the present model, including the theoretical uncertainty in the neutron EDM induced by the effective strong CP phase. 
The red shaded region indicates the projected CASPEr--Electric Phase-III sensitivity.}
  \label{fig:axion_photon}
\end{figure}

After removing the axion-dependent phases from the Yukawa sector, the color and electromagnetic anomalies of the PQ symmetry induce
\begin{align}
	\mathcal{L}_{\rm eff}^{\rm anom}
	=&~
	\frac{\alpha_s}{8\pi}
	\frac{a}{f_a}
	G_{\mu\nu}^a \widetilde{G}^{a\mu\nu}
	+
	\frac{\alpha_{\rm EM}}{8\pi}
	\frac{E}{N}
	\frac{a}{f_a}
	F_{\mu\nu}\widetilde{F}^{\mu\nu}.
\end{align}
Possible light components in the $45_H$ multiplet affect the prediction mainly through threshold corrections to GCU. 
The anomaly ratio $E/N$ remains unchanged, because these components are scalars. 
Such threshold effects shift the allowed $M_{\rm GUT}$ and $\alpha_{\rm GUT}$ region, and therefore change the predicted axion mass window.
At low energies, the axion-photon interaction is obtained by matching the effective theory to chiral perturbation theory. 
Including the model-independent contribution from axion--$\pi^0$ and axion--$\eta$ mixings, one obtains
\begin{align}
	\mathcal{L}_{a\gamma\gamma}
	=&
	-\frac{g_{a\gamma\gamma}}{4}
	a F_{\mu\nu}\widetilde{F}^{\mu\nu},
	\\
	g_{a\gamma\gamma}
	=&~
	\frac{\alpha_{\rm EM}}{2\pi f_a}
	\left(
	\frac{E}{N}
	-
	1.92(4)
	\right).
	\label{eq:axion-photon-coupling}
\end{align}
For the present model, we use $E/N=8/3$ in the numerical analysis. 
This value arises from the standard SU(5) embedding of the SM fermions into complete ${\bf 10}_F\oplus\overline{\bf 5}_F$ multiplets. 
The additional fields in $45_H$ are scalars and therefore do not contribute directly to the PQ anomaly.
In the left panel of Fig.\,\ref{fig:axion_photon}, we compare the predicted axion-photon coupling with the projected sensitivity of ABRACADABRA.

The gluonic anomaly term induces an effective strong CP phase,
\begin{align}
	\bar{\theta}_{\rm eff}
	=&~
	\frac{a}{f_a},
\end{align}
which induces an oscillating neutron EDM,
\begin{align}
	d_n(t)
	\simeq&~
	d_n^{(\bar{\theta})}\bar{\theta}_{\rm eff}
	\simeq
	2.4\times 10^{-16}
	\frac{a(t)}{f_a}\, e\,{\rm cm}.
\end{align}
Here, $a(t)$ denotes the coherently oscillating axion dark-matter background. Defining the axion-induced EDM coupling by
\begin{align}
	d_n(t)
	\equiv&~
	g_{aD}a(t),
\end{align}
we obtain
\begin{align}
	g_{aD}
	\simeq&~
	2.4\times 10^{-16}
	\frac{1}{f_a}\, e\,{\rm cm}.
\end{align}
Thus, the allowed range of $f_a$, determined by the GUT scale parameters, restricts the predicted range of the axion-induced EDM coupling $g_{aD}$.
CASPEr--Electric searches for spin precession induced by an oscillating EDM in an external electric field. 
Hence, the GUT-selected axion parameter region that determines $m_a$ and $g_{a\gamma\gamma}$ also predicts a definite region in the $m_a$--$g_{aD}$ plane.

We present the predicted axion parameter region in Fig.\,\ref{fig:axion_photon}.
In the present model, the allowed axion window is not fixed by the axion sector alone.
The GCU condition, the proton decay bounds, and the scalar-threshold spectrum associated with the $45_H$ multiplet determine the allowed values of $M_{\rm GUT}$ and $\alpha_{\rm GUT}$.
These correlations determine the axion window shown in Fig.\,\ref{fig:axion_photon}.

The charged-lepton proton decay constraint depends on the flavor parameter $V_e$.
Therefore, changing $V_e$ changes the lower edge of the allowed $M_{\rm GUT}$ region.
Since the GCU condition correlates $M_{\rm GUT}$ with the scalar threshold $M_{S_8}$, this also changes the allowed threshold region and, consequently, the allowed axion mass range.
For the representative values of $V_e$, we obtain
\begin{align}
	V_e=0.1:~
	1.88~{\rm neV}
	\leq&~ m_a
	\leq 12.1~{\rm neV},
	\\
	V_e=1.0:~
	1.94~{\rm neV}
	\leq&~ m_a
	\leq 3.84~{\rm neV},
	\\
	V_e=2.2:~
	1.96~{\rm neV}
	\leq&~ m_a
	\leq 2.56~{\rm neV}.
\end{align}
In Fig.\,\ref{fig:axion_photon}, we show the representative case $V_e=1.0$.

The left panel shows the prediction in the $m_a$--$g_{a\gamma\gamma}$ plane.
The allowed GUT parameter region fixes the range of $m_a$ and, together with $E/N=8/3$, determines the corresponding axion-photon coupling.
The prediction therefore appears as a line in the $m_a$--$g_{a\gamma\gamma}$ plane.

We compare this prediction with the projected sensitivity of ABRACADABRA under the benchmark assumptions used for the experimental reach~\cite{FileviezPerez:2019ssfa}.
The Phase-III broadband reach is evaluated for a magnetic field of $5~{\rm T}$ and a volume of $100~{\rm m}^3$, including only the irreducible noise contribution.
For the resonant Phase-III reach, the SQUID noise is assumed to be subdominant to the thermal noise.
The part of the predicted line covered by these sensitivities indicates the region in which future axion-photon coupling searches can test the model.

The right panel shows the corresponding prediction in the $m_a$--$g_{aD}$ plane.
An oscillating axion dark-matter background induces an oscillating neutron EDM through the effective strong CP phase.
The same range of $f_a$ fixes the corresponding range of the axion-induced EDM coupling $g_{aD}$.
The blue band shows the prediction of the present model, with its width reflecting the theoretical uncertainty in the neutron EDM induced by the effective strong CP phase.

We also show the projected sensitivity of CASPEr--Electric Phase-III.
The red shaded region denotes the expected experimental reach.
The comparison demonstrates that oscillating EDM searches can probe part of the axion parameter region selected by gauge coupling unification and proton decay constraints.

\section{Summary}
\label{sec:conclusion}

In this paper, we studied proton decay and axion phenomenology in an extended SU(5) GUT with a 45-dimensional Higgs field and a PQ sector.
The 45-dimensional Higgs field modifies the minimal fermion-mass relation through its Yukawa interactions and generates scalar-threshold corrections relevant to gauge coupling unification. 
In this sense, the model addresses several shortcomings of the minimal SU(5) GUT within a common framework.

We showed that successful gauge coupling unification can be achieved by a hierarchical mass spectrum for the scalar components of $\Phi_{45}$. 
In particular, the color-octet scalar $S_8$, the color-triplet scalar $S_3$, and the color anti-triplet scalar $S_1$ substantially affect unification when their masses lie significantly below the other components of $\Phi_{45}$.
The resulting threshold corrections enlarge the viable parameter region and keep gauge coupling unification consistent with current experimental constraints.

We also analyzed proton decay mediated by the exchange of the color anti-triplet scalar $S_1$. 
We estimated the relevant flavor structure of the $S_1$ couplings from the Georgi--Jarlskog texture and used it to evaluate the scalar-mediated proton decay rates.
Using these couplings, we computed the partial decay widths and examined their dependence on the model parameters. 
We found that the $p\to K^+\overline{\nu}$ mode provides the most stringent constraint on $M_{S_1}$ in the parameter region considered. 
The correlations among different decay modes, especially $p\to K^+\overline{\nu}$ and $p\to \mu^+K^0$, show that proton decay probes not only the mass scale of $S_1$ but also the flavor structure of the model.

The same parameter region selected by gauge coupling unification and proton decay constraints also restricts the axion observables.
The scalar thresholds from $\Phi_{45}$ affect the axion prediction indirectly through their impact on gauge coupling unification.
The anomaly ratio $E/N$, however, remains unchanged because the additional components in $\Phi_{45}$ are scalars.
For the present SU(5) embedding, we used $E/N=8/3$, which fixes the axion-photon coupling after the standard low-energy chiral matching.

We therefore evaluated the resulting axion predictions in the $(m_a,g_{a\gamma\gamma})$ and $(m_a,g_{aD})$ planes.
The predicted axion-photon coupling can be compared with the projected sensitivity of ABRACADABRA Phase-III, while the axion-induced neutron EDM coupling can be tested by CASPEr--Electric.
These results show that the model correlates fermion-mass generation, gauge coupling unification, proton decay, and axion searches.
Future proton decay searches and axion experiments will therefore provide complementary probes of this extended SU(5) framework.

Our results are also complementary to recent studies of proton decay as a diagnostic of ultraviolet physics, including analyses of chirality structures in nucleon-decay operators~\cite{Hamaguchi:2024ewe} and proton decay patterns from generic higher-dimensional operators \cite{Kitano:2026fri}.

\section*{Acknowledgments}
This work is partially supported by Scientific Grants by the Ministry of Education, Culture, Sports, Science and Technology of Japan, No.\,23K03392 (NH).

\appendix
\section{Details of the $S_1$-mediated Proton Decay}
\label{app:S1_proton_decay}
In this appendix, we collect the technical ingredients used to estimate the $S_1$-mediated proton decay rates in Sec.\,\ref{sec:proton decay}.
The assumptions on the $S_3$ contribution, the $S_1^{(5)}$--$S_1^{(45)}$ mixing, the Georgi--Jarlskog texture, and the numerical inputs are given in Sec.\,\ref{sec:proton decay}; here we only show the flavor estimates and the full set of partial width formulae.

For the order-of-magnitude estimate of the flavor rotations, we use the hierarchical forms
\begin{align}
M_u
\sim&~
v
\begin{pmatrix}
0 & \lambda^6 & 0\\
\lambda^6 & 0 & \lambda^2\\
0 & \lambda^2 & 1
\end{pmatrix}
\sim
v
\begin{pmatrix}
1 & \lambda^2 & 0\\
\lambda^2 & 1 & \lambda^2\\
0 & \lambda^2 & 1
\end{pmatrix}
\begin{pmatrix}
\lambda^8 & 0 & 0\\
0 & \lambda^4 & 0\\
0 & 0 & 1
\end{pmatrix}
\begin{pmatrix}
1 & \lambda^2 & 0\\
\lambda^2 & 1 & \lambda^2\\
0 & \lambda^2 & 1
\end{pmatrix},
\label{eq:app_Mu_hierarchy}
\\
M_d
\sim
M_e
\sim &~
v
\begin{pmatrix}
0 & \lambda^3 & 0\\
\lambda^3 & \lambda^2 & 0\\
0 & 0 & 1
\end{pmatrix}
\sim 
v
\begin{pmatrix}
1 & \lambda & 0\\
\lambda & 1 & 0\\
0 & 0 & 1
\end{pmatrix}
\begin{pmatrix}
\lambda^4 & 0 & 0\\
0 & \lambda^2 & 0\\
0 & 0 & 1
\end{pmatrix}
\begin{pmatrix}
1 & \lambda & 0\\
\lambda & 1 & 0\\
0 & 0 & 1
\end{pmatrix}.
\label{eq:app_MdMe_hierarchy}
\end{align}
Here $\lambda\simeq 0.22$, and coefficients of order unity are omitted.
The matrices multiplying the diagonal eigenvalue matrices give the parametric size of the rotations from the flavor basis to the fermion mass basis.

Using that the 45-dimensional Higgs contribution is dominated by the second-generation entry, the coupling matrix in the flavor basis is estimated as
\begin{align}
Y_{45}^D
\sim&~
\frac{v}{v_{45}}
\begin{pmatrix}
0 & 0 & 0\\
0 & \lambda^2 & 0\\
0 & 0 & 0
\end{pmatrix}.
\end{align}
The rotations in Eqs.~\eqref{eq:app_Mu_hierarchy} and
\eqref{eq:app_MdMe_hierarchy} then give
\begin{align}
\bigl(Y_1^{DU}\bigr)^T
\sim&~
\frac{v}{v_{45}}
\begin{pmatrix}
1 & \lambda^2 & 0\\
\lambda^2 & 1 & \lambda^2\\
0 & \lambda^2 & 1
\end{pmatrix}
\begin{pmatrix}
0 & 0 & 0\\
0 & \lambda^2 & 0\\
0 & 0 & 0
\end{pmatrix}
\begin{pmatrix}
1 & \lambda & 0\\
\lambda & 1 & 0\\
0 & 0 & 1
\end{pmatrix}
\sim
\frac{v}{v_{45}}
\begin{pmatrix}
\lambda^5 & \lambda^4 & 0 \\
\lambda^3 & \lambda^2 & 0\\
\lambda^5 & \lambda^4 & 0
\end{pmatrix},
\label{eq:app_Y1DU}
\\
Y_1^{QL,\ell}
\sim&~
\frac{v}{v_{45}}
\begin{pmatrix}
1 & \lambda^2 & 0\\
\lambda^2 & 1 & \lambda^2\\
0 & \lambda^2 & 1
\end{pmatrix}
\begin{pmatrix}
0 & 0 & 0\\
0 & \lambda^2 & 0\\
0 & 0 & 0
\end{pmatrix}
\begin{pmatrix}
1 & \lambda & 0\\
\lambda & 1 & 0\\
0 & 0 & 1
\end{pmatrix}
\sim
\frac{v}{v_{45}}
\begin{pmatrix}
\lambda^5 & \lambda^4 & 0\\
\lambda^3 & \lambda^2 & 0\\
\lambda^5 & \lambda^4 & 0
\end{pmatrix},
\label{eq:app_Y1QLell}
\\
Y_1^{QL,\nu}
\sim&~
\frac{v}{v_{45}} \,
\begin{pmatrix}
1 & \lambda & 0\\
\lambda & 1 & 0\\
0 & 0 & 1
\end{pmatrix} \,\,
\begin{pmatrix}
0 & 0 & 0\\
0 & \lambda^2 & 0\\
0 & 0 & 0
\end{pmatrix} \,\,
\begin{pmatrix}
1 & \lambda & 0\\
\lambda & 1 & 0\\
0 & 0 & 1
\end{pmatrix} \,\,
\sim
\frac{v}{v_{45}}
\begin{pmatrix}
\lambda^4 & \lambda^3 & 0\\
\lambda^3 & \lambda^2 & 0\\
0 & 0 & 0
\end{pmatrix}.
\label{eq:app_Y1QLnu}
\end{align}
Here, $Y_1^{QL,\ell}$ and $Y_1^{QL,\nu}$ denote the charged-lepton and neutrino components obtained from the same $SU(2)_L$ invariant coupling after rotating to the fermion mass basis. 
These components should not be regarded as independent Yukawa matrices. 
The corresponding flavor factors are defined in Sec.\,\ref{sec:proton decay} by Eqs.~\eqref{eq:deltaL_lepton_def} and \eqref{eq:deltaL_neutrino_def}. 
Since the complex conjugation of $Y_1^{DU}$ does not affect the power counting, the flavor suppressions follow directly from the above estimates.

The full set of $S_1$-mediated partial widths used in Table~\ref{tab:proton decay} is given by
\begin{align}
\Gamma(p\to\mu^+K^0)
=&~
\frac{1}{64\pi}
\left(1-\frac{m_K^2}{m_p^2}\right)^2
\frac{m_p}{f_\pi^2}
(1+D-F)^2
\alpha_H^2 A_{RL}^2
\frac{|\delta_{2112}^{L,\ell}|^2}{M_{S_1}^4},
\label{ptomuk}
\\
\Gamma(p\to\mu^+\pi^0)
=&~
\frac{1}{64\pi}
\left(1-\frac{m_\pi^2}{m_p^2}\right)^2
\frac{m_p}{f_\pi^2}
\frac{1}{2}
(1+D+F)^2
\alpha_H^2 A_{RL}^2
\frac{|\delta_{1112}^{L,\ell}|^2}{M_{S_1}^4},
\label{ptomupi}
\\
\Gamma(p\to e^+K^0)
=&~
\frac{1}{64\pi}
\left(1-\frac{m_K^2}{m_p^2}\right)^2
\frac{m_p}{f_\pi^2}
(1+D-F)^2
\alpha_H^2 A_{RL}^2
\frac{|\delta_{2111}^{L,\ell}|^2}{M_{S_1}^4},
\label{ptoek}
\\
\Gamma(p\to e^+\pi^0)
=&~
\frac{1}{64\pi}
\left(1-\frac{m_\pi^2}{m_p^2}\right)^2
\frac{m_p}{f_\pi^2}
\frac{1}{2}
(1+D+F)^2
\alpha_H^2 A_{RL}^2
\frac{|\delta_{1111}^{L,\ell}|^2}{M_{S_1}^4},
\label{ptoepi}
\\
\Gamma(p\to K^+\bar{\nu})
=&~
\frac{1}{64\pi}
\left(1-\frac{m_K^2}{m_p^2}\right)^2
\frac{m_p}{f_\pi^2}
\alpha_H^2 A_{RL}^2
\frac{1}{M_{S_1}^4}
\sum_{i=1}^{3}
\left|
\delta_{211i}^{L,\nu}
\frac{2D}{3}
+
\delta_{112i}^{L,\nu}
\left(1+\frac{D}{3}+F\right)
\right|^2,
\label{ptonuk}
\\
\Gamma(p\to \pi^+\bar{\nu})
=&~
\frac{1}{64\pi}
\left(1-\frac{m_\pi^2}{m_p^2}\right)^2
\frac{m_p}{f_\pi^2}
(1+D+F)^2
\alpha_H^2 A_{RL}^2
\frac{1}{M_{S_1}^4}
\sum_{i=1}^{3}
|\delta_{111i}^{L,\nu}|^2 .
\label{ptonupi}
\end{align}
Here, $m_p$, $m_K$, and $m_\pi$ denote the masses of the proton, kaon, and pion, respectively. 
The parameter $f$ denotes the pion decay constant, $D$ and $F$ are parameters of the baryon chiral Lagrangian, $\alpha_H$ is the hadronic form factor, and $A_{RL}$ accounts for renormalization-group evolutions of the dimension-six operators \cite{ClaudsonWiseHall1982,ChadhaDaniel1983,Nath:2006ut}.
For the hadron masses, we use the values in the Particle Data Group~\cite{ParticleDataGroup:2024cfk}. 
We take $D=0.80$, $F=0.46$, and $f_\pi=0.093~{\rm GeV}$. 
For the hadronic form factor, we use the lattice result $\alpha_H(2~{\rm GeV})=-\beta_H(2~{\rm GeV})=-0.0144~{\rm GeV}^3$ \cite{Aoki:2017puj}. 
Furthermore, the RGE of the dimension-six operators contributing to proton decay is taken into account through the factor $A_{RL}$. 
This factor is calculated by solving the one-loop RGE for the effective coupling of the proton decay operator, yielding $A_{RL}=2.6$.


\printbibliography

\end{document}